\documentclass[pra,aps,superscriptaddress,twocolumn,letter,nopacs,longbibliography]{revtex4-2}

\usepackage{array,multirow,graphicx,color,amsmath,amsfonts,enumerate,amsthm,amssymb,mathtools,enumitem,thmtools,hyperref,subfigure,mathdots,enumitem,centernot,bm,soul,bbm,tikz,pgfplots}
\usepackage{enumitem}
\usepackage[capitalise, noabbrev]{cleveref}
\usetikzlibrary{arrows}
\pgfplotsset{compat=1.14}
\hypersetup{colorlinks=true,linkcolor=blue,citecolor=blue,urlcolor=blue}

\def\>{\rangle}
\def\<{\langle}
\newcommand{\bra}[1]{\langle {#1} |}
\newcommand{\ket}[1]{| {#1} \rangle}
 
\newcommand{\ketbra}[2]{\ensuremath{\left|#1\right\rangle\!\!\left\langle#2\right|}}

\newcommand{\tr}[1]{\mathrm{Tr}\left[ #1 \right]}

\definecolor{ppblue}{RGB}{46,117,182}
\definecolor{ppred}{RGB}{197, 90, 17}

\newcommand{\magic}{T}
\newcommand{\rob}[2]{R(#1 | #2)}
\newcommand{\qrom}[2]{\mathcal{R}(#1 | #2)}
\usepackage{enumitem}
\newcommand{\monotone}{\mathcal{M}}

\theoremstyle{plain}

\theoremstyle{definition}

\newcolumntype{C}[1]{>{\centering\arraybackslash}p{#1}}

\definecolor{tikzBlue}{rgb}{0.6941176470588235,0.7568627450980392,0.8588235294117647}
\definecolor{tikzOrange}{rgb}{0.9294117647058824,0.7647058823529411,0.49019607843137253}
\definecolor{tikzBlue2}{rgb}{0.462745098,0.504575163,0.57254902}
\definecolor{tikzOrange2}{rgb}{0.619607843,0.509803922,0.326797386}
\definecolor{tikzGray}{rgb}{0.7529411764705882,0.7529411764705882,0.7529411764705882}

\DeclareFontFamily{U}{mathb}{\hyphenchar\font45}
\DeclareFontShape{U}{mathb}{m}{n}{
	<-6> mathb5 <6-7> mathb6 <7-8> mathb7
	<8-9> mathb8 <9-10> mathb9
	<10-12> mathb10 <12-> mathb12
}{}
\DeclareSymbolFont{mathb}{U}{mathb}{m}{n}
\DeclareMathSymbol{\llcurly}{\mathrel}{mathb}{"CE}
\DeclareMathSymbol{\ggcurly}{\mathrel}{mathb}{"CF}

\begin{document}

\title{Error mitigation and quantum-assisted simulation in the error corrected regime}
\author{M. Lostaglio}
\affiliation{QuTech, Delft University of Technology, P.O. Box 5046, 2600 GA Delft, The Netherlands}
\affiliation{Korteweg-de Vries Institute for Mathematics and QuSoft, University of Amsterdam, The Netherlands}
\author{A. Ciani}
\affiliation{QuTech, Delft University of Technology, P.O. Box 5046, 2600 GA Delft, The Netherlands}
 \affiliation{JARA Institute for Quantum Information, Forschungszentrum J\"ulich, D-52425 J\"ulich, Germany}

\begin{abstract}
A standard approach to quantum computing is based on the idea of promoting a classically simulable and fault-tolerant set of operations to a universal set by the addition of `magic' quantum states. In this context, we develop a general framework to discuss the value of the available, non-ideal magic resources, relative to those ideally required. We single out a quantity, the Quantum-assisted Robustness of Magic (QRoM), which measures the overhead of simulating the ideal resource with the non-ideal ones through quasiprobability-based methods. This extends error mitigation techniques, originally developed for Noisy Intermediate Scale Quantum (NISQ) devices, to the case where qubits are logically encoded. The QRoM shows how the addition of noisy magic resources allows one to boost classical quasiprobability simulations of a quantum circuit and enables the construction of explicit protocols, interpolating between classical simulation and an ideal quantum computer.

\end{abstract}

\maketitle

\emph{Introduction.}  
Large-scale quantum computing would allow us to solve computational problems that are intractable for classical computers. Due to the fragility of quantum information encoded in physical systems, quantum error correction will be a central component of these machines. However, several restrictions exist to the possibility of achieving a `universal' set of fault-tolerant quantum gates \cite{campbell2017roads, eastinknill}.   
%For quantum computing in the error corrected regime, resources are those computational elements that are not fault-tolerant and, as such, require special effort to be realized. 
 In a standard setting, the \emph{stabilizer operations} (which involve computational basis preparation and measurement, Clifford unitaries, partial trace and classical randomness \cite{AaronsonGottesman, seddonCampbell}) are fault-tolerant and hence `free'. Curiously, the Gottesman-Knill theorem tells us that these operations can be simulated efficiently on a classical computer~\cite{Gottesman:1998hu, AaronsonGottesman}. Stabilizer operations can be promoted to universality by injecting \emph{magic states}, which are thus a  `resource' for quantum computation. 
 Magic states may come at a limited rate, since their fault-tolerant preparation involves complex distillation schemes~\cite{bravyiKitaev, bravyiHaah, Fowler2013, meier2013, litinski2019, ogormancampbell2017}. Also, much of the residual noise in an error-corrected computation can originate from these elements.

In this work we go beyond the dichotomy free/resourceful and think in terms of the resource content of an  ideal quantum resource \emph{relative} to an available one.
We formulate questions such as ``How valuable is an ideal magic state, compared to its noisy version, as a function of the noise level?'' or  ``How valuable are the $r$ available magic states, compared to the $t>r$ ideally needed?''. Our notion of relative value stems from the overhead of  simulating ideal resources with non-ideal ones through quasiprobability based error mitigation \cite{temmeBravyi, kandala2019error, endobenjamin, takagi2020, endo2021}. As such, it is endowed with a clear operational significance. 

Quasiprobability-based error mitigation is a technique that allows one to remove bias from the outcome probabilities of a measurement by expressing the ideal circuit elements as linear combinations of non-ideal ones. The coefficients of the decomposition define a quasiprobability, since they are real and sum to one. By sampling from the absolute value of this quasiprobability
and performing the corresponding non-ideal operations, one can remove the bias at the price of a sampling overhead. 

We extend these ideas from Noisy Intermediate-Scale Quantum devices (NISQ) \cite{temmeBravyi, kandala2019error, endobenjamin, takagi2020, endo2021} to fault-tolerant quantum computing. In this context, we single out a measure, the Quantum-assisted Robustness of Magic (QRoM), which loosely speaking is a distance of the ideal elements relative to the available ones. The QRoM provides a unified setting to analyse several central computational and simulation tasks, investigated separately in the literature:

\emph{(1) Classical simulation:} classically sample from the outcome probabilities of an ideal quantum circuit with $t$ magic $T$-states \mbox{$\ket{T} = (\ket{0}+e^{i \pi/4}\ket{1})/\sqrt{2}$} as input ($t$ is known as the $T$-count). Having no quantum resources at hand, the QRoM coincides with the Robustness of Magic (RoM), a known measure of the overhead of quasiprobability-based classical simulation \cite{howardCampbell, heinrichGross2019, seddonCampbell, hakkakufujii2020}.

\emph{(2) Error mitigation in the quantum error corrected regime:} given a quantum circuit involving $t$ ideal $T$-states, obtain the average of measurement outcomes using a circuit with $t$ noisy $T$-states. The overhead is measured by the QRoM \footnote{A complementary approach analyzing error mitigation of decoding and approximating errors in the fault-tolerant regime has been discussed in Ref.~\cite{suzuki2020quantum}}.

\emph{(3) Quantum-Assisted Simulation:} obtain the average of measurement outcomes of an ideal quantum circuit involving $t$ ideal $T$-states, given that we can inject only $r < t$ noisy $T$-states. This task can be seen as an intermediate scenario between classical simulation ($r=0$) and error mitigation $(r=t)$, where some quantum resources are available but fewer than what is needed.  
The residual overhead is quantified by  the QRoM.

By analytically or numerically solving optimization problems for the QRoM, we construct  quasiprobability-based error mitigation and quantum-assisted simulation algorithms in the error corrected regime, and analyze how the addition of quantum resources gradually interpolates between a classical simulation and an ideal quantum computation.

\emph{Quantum-assisted Robustness of Magic.}
The QRoM is a generalization of the RoM and formalizes the idea of robustness of an ideal quantum computational resource relative to the (fewer, noisier) available ones. Consider a quantum computer where magic states are injected by means of $t$ ancilla qubits. A resource state $\sigma_r$ on \mbox{$r\leq t$} qubits is given,  e.g., $T$-states resulting from a number of distillation rounds. Due to practical limitations, these states are in general fewer and noisier than required by the ideal circuit. 

In broad generality, one can denote by $\mathcal{Q}_t(\sigma_r)$ all the $t$-qubit states achievable from the resource $\sigma_r$ by means of stabilizer operations.
\begin{figure}
    \centering
    \includegraphics[width=0.25\textwidth]{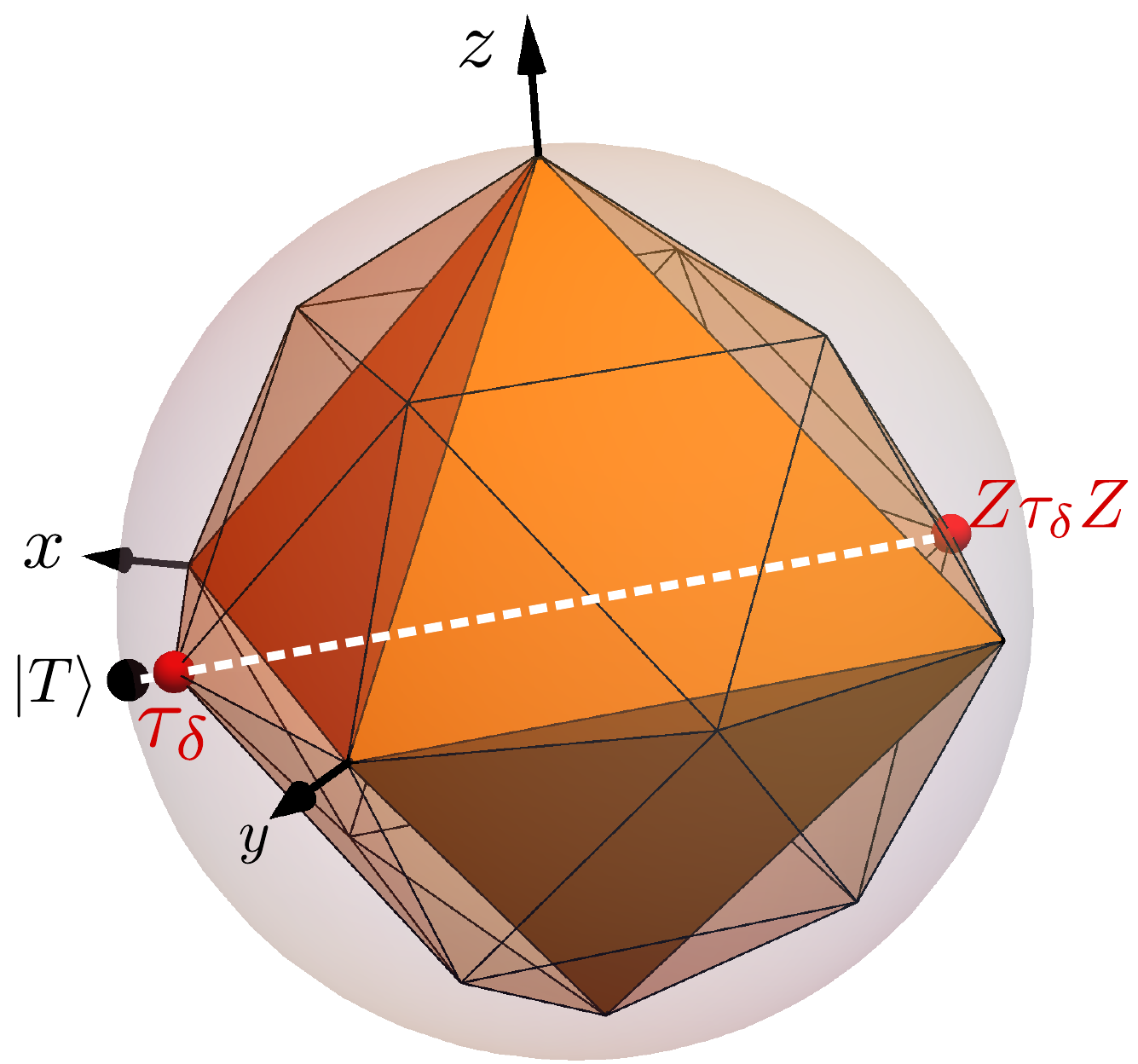}
    \caption{\emph{Relative robustness for a single qubit in the Bloch sphere.} The black dot represents the pure magic state $\ket{T}$. The orange, full octahedron represents the convex hull of stabilizer states, while the orange, shaded region, is the set $\mathcal{Q}_1(\tau_\delta)$ representing all states achievable from the available resource (a noisy $T$-state $\tau_{\delta} = (1-\delta) \ketbra{T}{T} + \delta I/2$ with $\delta =0.1$) by stabilizer operations.  Geometrically, the relative robustness is the minimum mixing needed to bring the $T$-state inside $\mathcal{Q}_1(\tau_\delta)$ (in this case, mixing with $Z \tau_{\delta} Z$).}
    \label{fig:polytope}
\end{figure}
The \emph{relative} robustness of a $t$-qubit ideal resource $\rho_t$ with respect to the available resource $\sigma_r$ is
\small
\begin{equation}
\label{eq:qrom}
\rob{\rho_t}{\sigma_r} = \min \left\{ s\geq 0\biggr| \frac{\rho_t + s \eta}{1+s} \in \mathcal{Q}_t(\sigma_r), \; \eta \in  \mathcal{Q}_t(\sigma_r) \right\}.
\end{equation}
\normalsize
The relative robustness represents the minimum amount of mixing between $\rho_t$ and a state in $\mathcal{Q}_t(\sigma_r)$ such that the resulting state is also in $\mathcal{Q}_t(\sigma_r)$.
Since all pure stabilizer states on $t$ qubits can be generated by applying $t$-qubit Clifford unitaries to $\ket{0}^{\otimes t}$ \cite{AaronsonGottesman}, $R(\rho_t|\sigma_r)$ is the robustness relative to a set given by  stabilizer states \emph{augmented} by the available resources, as illustrated in Fig.~\ref{fig:polytope}.

The QRoM is the negativity required to decompose the ideal resource $\rho_t$ in terms of the available ones in $\mathcal{Q}_t(\sigma_r)$:
\begin{multline}
\label{eq:quacnegativity}
\qrom{\rho_t}{\sigma_r}  = \min \biggl\{ \sum_x |q_x| \biggr| \rho_t = \sum_{x} q_{x} \eta_x, \eta_x \in \mathcal{Q}_t(\sigma_r)  \biggr\}.
\end{multline}

Note that $q_x \in \mathbb{R}$ and $\sum_x q_x =1$, so $q_x$ defines a quasiprobability. The two quantities just defined are closely related, as they satisfy~$ \qrom{\rho_t}{\sigma_r}  =1 + 2  \rob{\rho_t}{\sigma_r}$~(see Appendix \ref{app:robustnessandnegativity})
When $\sigma_r$ can be written as a mixture of stabilizer states, the QRoM reduces to the well-known RoM of Ref.~\cite{howardCampbell} (denoted by $\mathcal{R}(\rho_t)$), which measures the overhead to classically simulate the resource $\rho_t$ via the Gottesman-Knill theorem. We will show that $\qrom{\rho_t}{\sigma_r}$ has an analogue interpretation in the presence of the quantum resource $\sigma_r$. The QRoM has desirable properties of a resource theoretical measure \cite{Note3}, similar to those of the Wigner negativity  \cite{takagiQuntao, albarelli2019, veitch2014resource, marieisert}: faithfulness, monotonicity, convexity (see Appendix \ref{app:propquac}).

 In general, \mbox{$\qrom{\rho_t}{\sigma_r}\leq \mathcal{R} (\rho_t)$}. As we shall see, the QRoM makes quantitative the intuition that the available quantum resources decrease the negativity and hence the computational overhead. The QRoM is an operationally motivated concept suitable to study classical simulation, quantum-assisted simulation and error mitigation in the error corrected regime under a unified umbrella.

\emph{Error mitigation in the quantum error corrected regime.} 
The first task where the QRoM plays a role is error mitigation.  We want to perform a quantum computation on $n$ data qubits, which can be assumed to be initialized in state $\ket{0}^{\otimes n}$. The circuit involves Clifford unitaries, typically taken from a fundamental gate set, and $T$-gates:
\begin{equation}
T = \ketbra{0}{0} + e^{i \pi/4} \ketbra{1}{1}.
\end{equation}
The latter promote the computation to universality~\cite{nielsenChuang}.  Since we assume our qubits to be logically encoded in a suitable quantum error correcting code, we take all Clifford operations to be perfect (this assumption will be relaxed later). Each one of the $t$ $T$-gates is realized via a gate teleportation gadget involving a $T$-state~$\ket{T}$~\footnote{See Ref.~\cite{ campbell2017roads}. For a reminder, see Appendix \ref{app:reminder}.}. In practice, $T$-states will be noisy. While our approach is general, for simplicity we focus on noisy magic states with standard form
\begin{equation}
    \tau_\delta = (1-\delta) \tau + \delta I/2,
\end{equation}
where $\tau = \ketbra{T}{T}$, $I$ is the identity matrix and $\delta \in [0, 1]$ is the noise level~\footnote{If $\ket{T}$ is affected by a different noise, it can be brought into this standard form by applying the identity or the Clifford unitary $ \exp[-i \pi/4] S X$ with probability $1/2$, as also remarked in Ref.~\cite{bravyiHaah}. Here $S$ is the phase gate and $X$ the Pauli $x$ unitary.}. The quantum computation terminates with the measurement of a Pauli operator $P$. We  want to estimate its average $\langle P \rangle$ in the final state of the ideal circuit.

Since $T$-states are noisy, the measured average cannot be expected to be an unbiased estimator of $\langle P \rangle$. In order to cancel the bias via error mitigation, first we find a decomposition of the form in Eq.~\eqref{eq:quacnegativity}. Ideally we wish to find an optimal decomposition according to Eq.~\eqref{eq:quacnegativity} (with $\rho_t = \tau^{\otimes t}$ and $\sigma_{t} = \tau^{\otimes t}_{\delta}$), but
in practice the decomposition need not be optimal. Setting $\|q\|_1= \sum_x |q_x|$, the algorithm works as follows:
\begin{enumerate}
\item Sample $x$ with probability $|q_{x}|/\|q\|_1$.
\item Run the quantum circuit with input $\eta_x \otimes \ketbra{0}{0}^{\otimes n}$ \footnote{Perhaps one cannot directly sample $\eta_x$. In fact, according to the definition of $\mathcal{Q}_t(\sigma_r)$, we may be able to only prepare some states $\phi_i$ such that $\eta_x = \sum_{i} h^{(x)}_i \phi_i$ for a probability $\{h^{(x)}_i\}_i$. If that's the case, sample $\phi_i$ with probability $h^{(x)}_i$. An application of Hoeffding's inequality shows that this sampling has the same overhead as the original one.}.
\item Measure the Pauli operator $P$, getting outcome $p=-1$ or $1$.
\item Output $o=p \lVert q \rVert_1 \mathrm{sign}(q_{x})$.
\end{enumerate}
Sampling $M$ times and taking the arithmetic average of the outputs, one obtains an unbiased estimator for $\langle P \rangle$. By Hoeffding's inequality \cite{hoeffding1963}, after $M \geq 2\ln(2/\epsilon) \lVert q \rVert_1^2/ \Delta^2 $ runs the estimate is within additive error $\Delta$ of $\langle P \rangle$ with probability $1-\epsilon$ (see Appendix \ref{app:overheadandQRoM}). Hence, this protocol cancels the error on $\langle P \rangle$ from the noisy $T$-states at the price of a sampling overhead. By using an optimal decomposition,
\begin{equation}
    M = 2\ln (2/\epsilon) \qrom{\tau^{\otimes t}}{\tau^{\otimes t}_{\delta}}^2/ \Delta^2
\end{equation}
suffices. Since for an ideal quantum computer one needs \mbox{$M_{\textrm{ideal}} = 2\ln(2/\epsilon)/ \Delta^2$} runs to obtain the same guarantee, the QRoM squared quantifies the overhead.

Finding the optimal decomposition in Eq.~\eqref{eq:quacnegativity} is a convex optimization problem~\cite{boyd}  whose size scales super-exponentially with the $T$-count~\footnote{It is possible to alleviate this problem by exploiting the symmetries of the state $\tau^{\otimes t}$ \cite{heinrichGross2019}.}. While such an approach is unscalable, we obtain upper and lower bounds
\begin{equation}
\label{eq:bounds}
   \frac{\mathcal{M}(\tau^{\otimes t})}{\mathcal{M}(\tau^{\otimes t}_\delta)} \leq \qrom{\tau^{\otimes t}}{\tau^{\otimes t}_{\delta}} \leq [\qrom{\tau^{\otimes k}}{\tau^{\otimes k}_{\delta}}]^{t/k}.
\end{equation}

 Lower bounds can be derived from any convex magic monotone $\mathcal{M}$, see Appendix \ref{app:lowerbound}. Taking $\mathcal{M}$ to be the dyadic negativity \cite{seddon2020}, we show  that $\qrom{\tau^{\otimes t}}{\tau^{\otimes t}_{\delta}}$ grows exponentially with the $T$-count $t$ for any $\delta \in  (0, 1]$:
\begin{equation}
\nonumber
   \mathcal{R}(\tau^{\otimes t} | \tau^{\otimes t}_{\delta}) \geq  \min \left\{ \frac{1}{(1-\delta/2)^t}, [2(2-\sqrt{2})]^t \right\}.
\end{equation}
The upper bounds are based on the submultiplicativity of the QRoM. Taking $t$ to be divisible by $k$, $\tau^{\otimes t} = \tau^{\otimes k} \otimes \dots \otimes \tau^{\otimes k}$, decompositions for $\tau^{\otimes k}$ give (sub-optimal) block decompositions for $\tau^{\otimes t}$. These can be used in practical error mitigation protocols. 

Let us start with $k=1$. The QRoM is
\begin{equation}\label{eq:neg1t1r}
\qrom{\tau}{\tau_{\delta}} = \begin{cases}
\sqrt{2} = \mathcal{R}(\tau), \quad &\delta > \delta_{\mathrm{th}}, \\
\frac{1}{1- \delta}, \quad &\delta \le \delta_{\mathrm{th}}.
\end{cases}
\end{equation}
Above a noise threshold $\delta_{\mathrm{th}}= 1- \sqrt{2}/2 \approx 0.293$ the RoM is recovered~\cite{howardCampbell}.
The optimal decomposition for $\delta \leq \delta_{\textrm{th}}$ is instead
\begin{equation}\label{eq:silly2}
\tau = \frac{1 - \delta/2}{1-\delta} \tau_{\delta} - \frac{\delta/2}{1- \delta} Z \tau_{\delta} Z,
\end{equation}
where $Z$ is the Pauli $z$  matrix. The threshold corresponds to the point where the noisy magic state $\tau_{\delta}$ enters the set of stabilizer states (inner octahedron in Fig.~\ref{fig:polytope}). By tuning the noise, the QRoM interpolates between error mitigation and classical simulation. The latter is recovered at $\delta = \delta_{\textrm{th}}$, since at that point the noise is so large that the quantum resources can be classically simulated efficiently. 
 
\begin{figure}[t]
   \includegraphics[width=0.378 \textwidth]{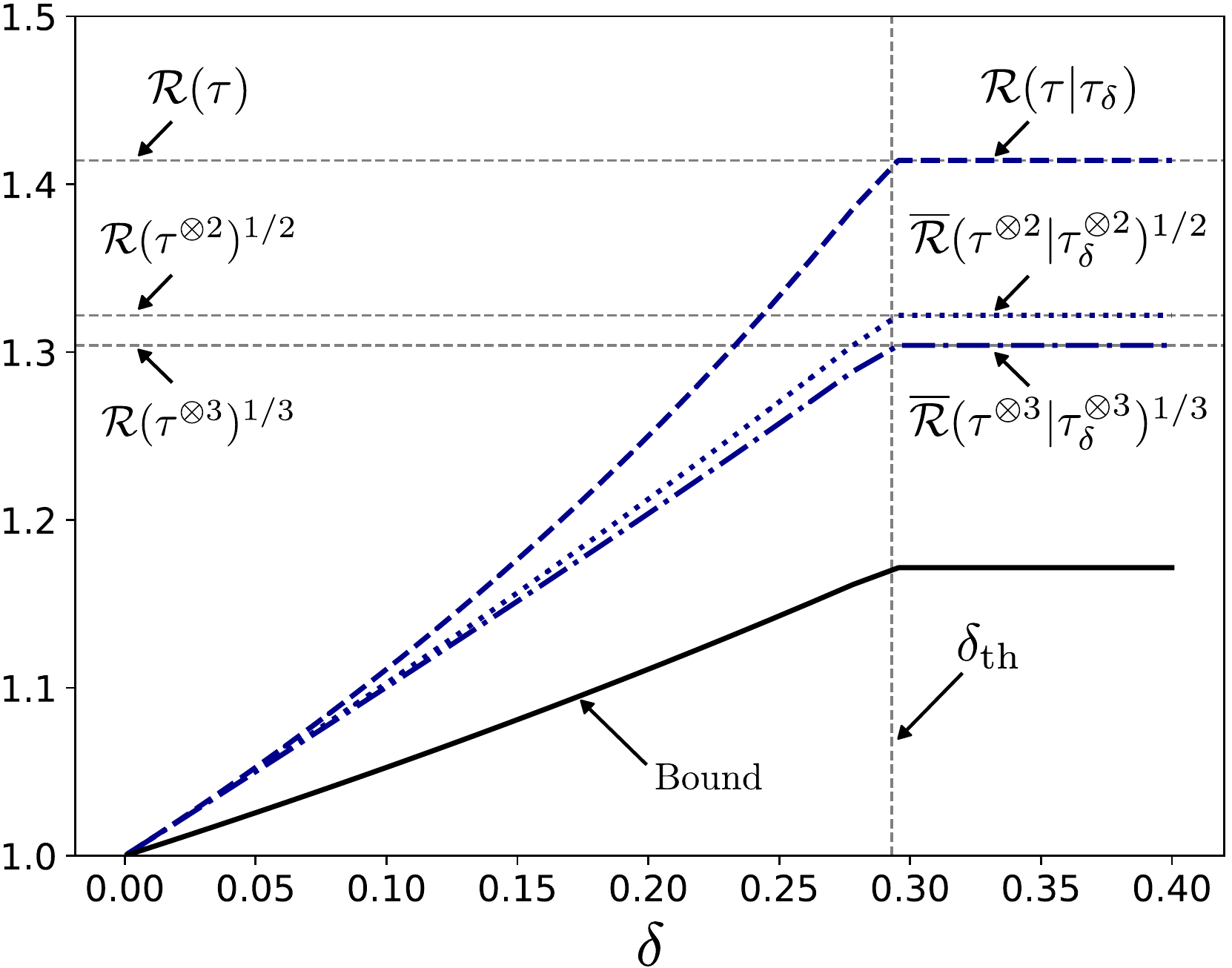}
    \caption{\emph{QRoM as a function of noise.} $k$-th root of the QRoM of $\tau^{\otimes k} $ with respect to $\tau_{\delta}^{\otimes k}$, as a function of the noise level $\delta$.  
    $\overline{\mathcal{R}}(\tau^{\otimes k}| \tau_\delta^{\otimes k})$ are upper bounds for the corresponding QRoM, based on analytical decompositions. At $\delta =\delta_{\mathrm{th}}$, the $k=2$ ($k=3$) upper bound recovers exactly (approximately) the RoM $\mathcal{R}(\tau^{\otimes k})$ of Ref.~\cite{howardCampbell}. The black solid line represents the bound $\mathcal{R}_{\textrm{bound}}(\tau^{\otimes t}| \tau_\delta^{\otimes t})^{1/t}$, which for small $\delta$ goes as $1+\delta/2$. While $k=2, 3$ outperform  $k=1$, the improvement is modest for small level of noise.} 
    \label{fig:error_mitigation}
\end{figure}

For $k=2,3$ the situation is similar, but computing the QRoM exactly is hard. In Appendix \ref{app:analytic} we obtain explicit decompositions whose performance is presented in Fig.~\ref{fig:error_mitigation}.

 In the small error regime, realistic for quantum error corrected setups $(\delta \leq 10^{-2})$, global decompositions only marginally outperform the single-qubit decomposition of Eq.~\eqref{eq:silly2}~\footnote{For example, the $k=3$ block decomposition error mitigates a quantum circuit with depolarizing noise $\delta =10^{-2}$ on each $T$-state with sampling overhead $1.0201^t$, while $k=1$ achieves $1.0203^t$. Formally, we conjecture $\qrom{\tau^{\otimes t}}{\tau_\delta^{\otimes t}} \approx (1 + \delta)^t$ as $\delta \rightarrow 0$.}.
Already at $\delta=10^{-2}$ and moderate overheads ($\sim 10^2$) one can error mitigate in regimes ($t \sim 230$) in which classical simulation is currently unfeasible even with state-of-the-art algorithms  \cite{bravyigosset2016, pashayankorzekwa2021, seddon2020}.

\emph{Error mitigation with noisy Cliffords.} So far we assumed that Clifford unitaries are ideal. In reality, they have a residual noise. In this situation it is more natural to perform error mitigation at the level of channels rather than states. Consider a simple error model where each $k=1,2$ qubit Clifford $\mathcal{U}^{(k)}$ is independently affected by depolarizing noise:
\begin{equation}
     \mathcal{U}^{(k)}_{\delta_c} = (1-\delta_c)  \mathcal{U}^{(k)} + \delta_c \mathcal{G}_k
\end{equation}
where $\mathcal{G}_k(\rho) = I/2^k$ for all $\rho$ and $\delta_c \in [0, 1]$ is the noise level on the Cliffords. The noise affects not only the Clifford unitaries on the data qubits, but also those involved in the gate teleportation gadget. The implementable $T$-gates, denoted by $\mathcal{T}_{\delta,\delta_c}$, are then noisy both due to $\delta>0$ and $\delta_c>0$. Consider an ideal circuit realizing a unitary $\mathcal{U}$ by a sequence of $n^{(1)}_c$ single-qubit Cliffords, $n^{(2)}_c$ two-qubit Cliffords and $t$ $T$-gates. To reproduce the expectation values $\langle P \rangle$ of this ideal circuit, find a quasiprobability decomposition $\mathcal{U} = \sum_{x} q_{x} \mathcal{A}_x$, with $\mathcal{A}_x$ available noisy channels (in this case, $\mathcal{A}_x = \mathcal{T}_{\delta,\delta_c}, \mathcal{U}^{(k)}_{\delta_c}$ and compositions thereof). The minimal negativity over all such decomposition is the ``channel QRoM'' $\qrom{\mathcal{U}}{\{\mathcal{A}_x\}}$.

Given any decomposition, a protocol cancelling the bias on $\langle P \rangle$ due to noisy gates is obtained following steps 1,3,4, of the previous algorithm, but replacing step 2 with
\begin{enumerate}
    \item[2'] Run the noisy quantum circuit $\mathcal{A}_x$. 
\end{enumerate}
As before, after $M \geq 2 \ln(2/\epsilon)\|q\|^2_1/\Delta^2$ runs the estimate is within error $\Delta$ of $\langle P \rangle $ with probability $1-\epsilon$. $\qrom{\mathcal{U}}{\{\mathcal{A}_x\}}^2$ is then the minimal sampling overhead. In Appendix \ref{app:noisycliffords} we obtain a (block) decomposition of $\mathcal{U}$ by separately finding a quasiprobability decomposition for $T$-gates, one and two-qubit Clifford gates.

For $\delta < \delta_{\mathrm{th}}$, the corresponding sampling overhead (which upper bounds $\qrom{\mathcal{U}}{\{\mathcal{A}_x\}}$) is

\small
\begin{equation}
\nonumber
   \left(\frac{2-\delta}{(1-\delta) (1-\delta_c)^2}-1\right)^{2 t} \left(\frac{1+\frac{\delta_c}{2}}{1-\delta_c}\right)^{2 n^{(1)}_c}\left(\frac{1+\frac{7 \delta_c}{8}}{1-\delta_c}\right)^{2 n^{(2)}_c} 
\end{equation}
\normalsize
For $\delta_c=0$ we recover $1/(1-\delta)^{2 t}$, as expected from Eq.~\eqref{eq:neg1t1r} \footnote{For the sake of comparison, setting $\delta =10^{-2}$ and \mbox{$\delta_c = 10^{-3}$} we get an overhead  $1.02845^t \times 1.00301^{n^{(1)}_c} \times 1.00376^{n^{(2)}_c}$.}. The $T$-gate overhead is rather close to the one found for ideal Clifford gates. 

\begin{figure}[t]
    \includegraphics[width=0.5 \textwidth]{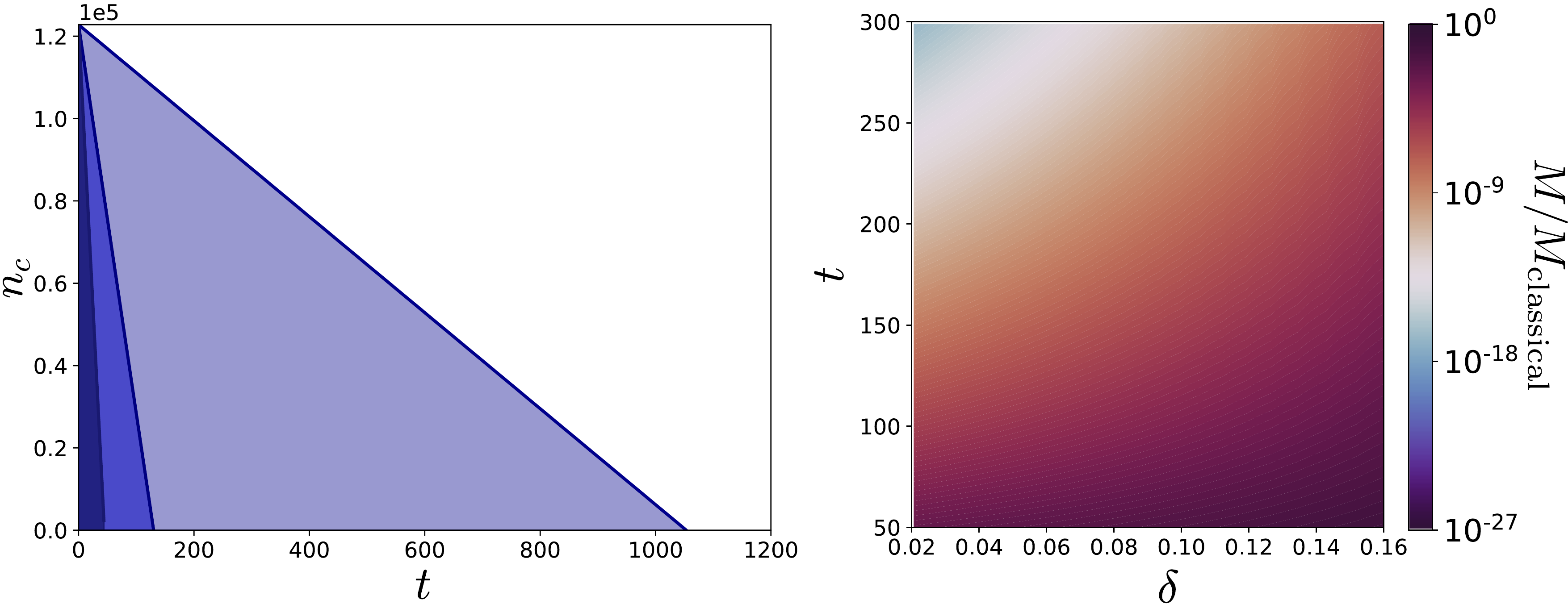}
    \caption{\emph{Left: Error mitigation and distillation.} Maximum number of $T$-gates ($t$) and Cliffords ($n_c$) that can be error mitigated with moderate overhead $(\leq 100)$ after $0$, $1$ and $2$ rounds of ideal Bravyi-Haah $14\mapsto 2$ distillation (from dark to lighter). The initial noise parameters are $\delta_c = 10^{-5}$ and $\delta=5 \times 10^{-2}$. We took the worst-case scenario where $n_c$ are all two-qubit Cliffords. \emph{Right: Classical vs quantum-assisted simulation.} Ratio between the quantum-assisted and classical number of samples $M/M_{\mathrm{classical}}$ needed to estimate the average value of a Pauli observable with given precision, as a function of the noise $\delta$ and the $T$-count $t$ of the circuit, for $t/r=3$ (inject a third of the required $T$-states).}
    \label{fig:quac}
    \label{fig:distillation}
\end{figure}

Magic state distillation protocols \cite{campbell2017roads}  can be combined with error mitigation, in order to decrease the initial $T$-state error. In Fig.~\ref{fig:distillation} we illustrate this by presenting the maximum number of Cliffords and $T$-gates whose noise can be mitigated with moderate overhead ($\leq 10^2$) after zero, one or two rounds of the Bravyi-Haah $14 \mapsto 2$ magic state distillation protocol \cite{bravyiHaah}. To highlight the performance of the error mitigation stage, we assumed an ideal distillation protocol. However, we show in Appendix \ref{app:mitigate_dist} that also the noise in the Clifford unitaries required in the distillation round can be error mitigated.

\emph{Quantum-Assisted Simulation.}  What happens when the $T$-count $t$ of the ideal quantum circuit exceeds the available resources? That is indeed a generic situation. Specifically, suppose we have at our disposal only $r<t$ noisy $T$-states and ideal Cliffords to simulate the circuit. We call this task \emph{quantum-assisted simulation}, as it interpolates between classical simulation (where $r=0$) and error mitigation (where $r=t$).

Injection of insufficient $T$-states, as well as noise on each $T$, both lead to bias on the expectation value $\langle P \rangle$ which can be corrected by quasiprobability methods. To do so, find a decomposition of the form in Eq.~\eqref{eq:quacnegativity} with $\rho_t = \tau^{\otimes t}$ and $\sigma_r = \tau^{\otimes r}_\delta$, with $r$ a fraction of $t$. Then, apply the algorithm steps 1-4 above. By construction, in step 2 of the protocol at most $r$ noisy magic states are injected, rather than the $t$ ideally required. Nevertheless, after taking
\begin{equation}
    M = 2\ln (2/\epsilon) \qrom{\tau^{\otimes t}}{\tau^{\otimes r}_{\delta}}^2/ \Delta^2
\end{equation}
samples, the output average will be within $\Delta$ of the ideal average with probability $1-\epsilon$. From
\begin{equation}
    \qrom{\tau^{\otimes t}}{\tau^{\otimes r}_{\delta}} \leq [\qrom{\tau^{\otimes t/r}}{\tau_{\delta}}]^{r},
\end{equation}
explicit protocols can obtained by decomposing each $t/r$ $T$-states using a single noisy $T$-state.

We find find analytical decompositions upper bounding the QRoM for $t/r =2$ and $t/r =3$ (see Appendix \ref{app:analytic}). The performance of the corresponding protocols is compared to the best known classical algorithms based on the RoM in Fig.~\ref{fig:quac}. For $\delta = 10^{-2}$ and $t/r=3$ (inject a third of the required $T$-states) we get a scaling overhead $1.45715^t$, compared to $1.667^t$ of the best classical algorithm given in Ref.~\cite{heinrichGross2019}. 
Even injecting a fraction of the required quantum resources leads to orders of magnitudes improvements over the classical protocol.

The previous protocol injects fewer magic states at the price of a sampling overhead. Nevertheless, it still requires an $n$-qubit quantum computer. However, suppose we only have $r<n$ noisy qubits and we want to use them to assist a classical computer to estimate outcome probabilities of an ideal circuit with $T$-count $t= k r$ ($k=2,3,\dots$). 
We can proceed as follows. If $P = \Pi_1 - \Pi_{-1}$, in steps 2-3 of the previous algorithm we need to sample the measurement outcome $p=0,1$ with probability
	\begin{equation}
	\label{eq:sample}	
\tr{ \Pi_p \mathcal{A}  \left(\mathcal{U}_a (\tau^{\otimes s}_\delta \otimes \ketbra{0}{0}^{t-s}) \otimes \ketbra{0}{0}^{\otimes n}\right)},
	\end{equation} 
where $\mathcal{A}$ is some (adaptive) Clifford circuit and $s \leq r$. To sample from the above, we make use of the extended Gottesman-Knill theorem~ \cite{yoganathan2019quantum} (also see \cite{bravyi2016trading, bravyigosset2016, pashayankorzekwa2021}). Given a description of $\mathcal{A}$ we classically obtain in \mbox{$\textrm{poly}(n+t)$} time a description of a new adaptive Clifford circuit $\overline{\mathcal{A}}$ involving at most $s$ measurements. Then from a sample of $\tr{ \Pi^{(p)} \overline{\mathcal{A}} (\tau_{\delta}^{\otimes s})}$
and a $\textrm{poly}(n+t)$ computation one can obtain a sample of Eq.~\eqref{eq:sample}. Steps 2-3 have now been reduced to a task that can be performed on the $s$-qubit noisy quantum computer. The number of required samples is again quantified by the QRoM $\mathcal{R}(\tau^{\otimes t/r}| \tau_\delta)^r$.

\emph{Discussion.} Our analysis led to practical protocols for error mitigation in currently classical intractable regimes and quantum-assisted simulation outperforming the best known quasiprobability-based algorithms. These proposals appear relevant for a regime in which full magic state distillation is unavailable. Alternative QRoMs can also defined, e.g., by restricting stabilizer operations to account for specific architecture limitations. In this regard it should be also noted that our work, like all quasiprobability-based error cancellation methods, is limited by the requirement of well-characterized noise. We suggest that similar approaches can be envisioned to boost alternative simulation methods based on the stabilizer rank ~\cite{bravyismolin2016, bravyigosset2016, bravyicampbell2019, seddon2020, pashayankorzekwa2021} or on generalized Wigner functions \cite{raussendorf2020, pashayan2015}.

\bigskip

\textbf{Acknowledgements.}
Our work was supported by ERC grant EQEC No. 682726. AC acknowledges funding from  the Deutsche Forschungsgemeinschaft (DFG, German Research Foundation) under Germany's Excellence Strategy – Cluster of Excellence Matter and Light for Quantum Computing (ML4Q) EXC 2004/1 – 390534769. We thank B. Terhal for useful discussions.

\textbf{Note.} During the preparation of this manuscript, we became aware of an independent effort to use error mitigation for universal quantum computing via encoded Clifford+T circuits \cite{piveteau2021error}.

\appendix

\section{Robustness and negativity}
\label{app:robustnessandnegativity}

Here we prove that $\qrom{\rho_t}{\sigma_r} = 1 + 2 \rob{\rho_t}{\sigma_r}$. 

\begin{proof}
The proof follows using standard techniques. First we prove $\qrom{\rho_t}{\sigma_r}   \leq 1 + 2 \rob{\rho_t}{\sigma_r}$. Let $\xi$, $\eta$ be elements of $\mathcal{Q}_t(\sigma_r)$ such that
\begin{equation}
    \xi = \frac{\rho_t + \rob{\rho_t}{\sigma_r} \eta}{1+ \rob{\rho_t}{\sigma_r}}. 
\end{equation}
Then $\rho_t = (1+ \rob{\rho_t}{\sigma_r} )\xi - \rob{\rho_t}{\sigma_r} \eta$. By decomposing $\eta$ and $\xi$ in terms of the states in $\mathcal{Q}_t(\sigma_r)$, we obtain a decomposition of the form in Eq. 2 in the main text with $\|q\|_1 = 1 + 2 \rob{\rho_t}{\sigma_r}$. Since $\|q\|_1 \geq \qrom{\rho_t}{\sigma_r}$, it follows that $\qrom{\rho_t}{\sigma_r}\leq 1 + 2 \rob{\rho_t}{\sigma_r} $.

Now we prove $\qrom{\rho_t}{\sigma_r} \geq 1+ 2 R(\tau_t|\rho_r)$. Consider an optimal decomposition
\begin{equation}
    \rho_t = \sum_x q_x \eta_x, 
\end{equation}
where $\eta_x \in \mathcal{Q}_t(\sigma_r)$ and $\|q\|_1 = \qrom{\rho_t}{\sigma_r}$. Then
\begin{equation}
   \rho_t = \sum_{x|q_x \geq 0} q_x \eta_x - \sum_{x|q_x <0} |q_x| \eta_x = (1+s) \xi - s \eta,
\end{equation}
where $s = \sum_{x|q_x <0} |q_x|$ and
\begin{equation}
     \xi =  \frac{1}{1+ s} \sum_{x|q_x \geq 0} q_x \eta_x, \quad \eta = \frac{1}{s}  \sum_{x|q_x <0} |q_x| \eta_x. 
\end{equation}
Note that $\xi,\eta \in \mathcal{Q}_t(\sigma_r)$ since $\mathcal{Q}_t(\sigma_r)$ is a convex set. It follows that
\begin{equation}
    \frac{\rho_t + s \eta}{1+s} \in \mathcal{Q}_t(\sigma_r).
\end{equation}
Then $\rob{\rho_t}{\sigma_r}\leq s$. However, $\qrom{\rho_t}{\sigma_r}= 1+2s$, so we conclude $\qrom{\rho_t}{\sigma_r} \geq 1+ 2 \rob{\rho_t}{\sigma_r}$.
\end{proof}

\section{Properties of the QRoM}
\label{app:propquac}
In this appendix we show for completeness some properties of the QRoM $\qrom{\rho_t}{\sigma_{r}}$ defined in Eq. 2 of the main text. Most of the properties and their proofs are straightforward generalizations of those of the RoM $\mathcal{R}(\rho_t)$ that can be found in Ref. \cite{howardCampbell}.
\begin{enumerate}[label=P\arabic*]
\item Faithfulness: $\qrom{\rho_t}{\sigma_r} = 1$ if and only if \mbox{$\rho_t \in \mathcal{Q}_t(\sigma_r)$}.
\item Sub-multiplicativity with respect to both arguments: if $r_{1, 2} \le t_{1,2}$
\begin{equation}
    \qrom{\rho_{t_1} \otimes \rho_{t_2}}{\sigma_{r_1} \otimes \sigma_{r_2}} \le \qrom{\rho_{t_1}}{ \sigma_{r_1}} \qrom{\rho_{t_2}}{\sigma_{r_2}}.
\end{equation}  
\item Monotonicity: let $\mathcal{E}$ be a stabilizer operation. Then \begin{align}
    \qrom{\mathcal{E}(\rho_t)}{\mathcal{E}(\sigma_r)} \le \qrom{\rho_t}{\sigma_r}, \\ 
    \qrom{\rho_t}{\mathcal{E}(\sigma_r)} \ge \qrom{\rho_t}{\sigma_r}.
\end{align}

\item Invariance under Cliffords: let $\mathcal{U}$ be a Clifford unitary CPTP map. Then $\qrom{\mathcal{U}(\rho_t)}{\sigma_r} = \qrom{\rho_t}{\sigma_r}$  and $\qrom{\rho_t}{\mathcal{U}(\sigma_r)} = \qrom{\rho_t}{\sigma_r}$.

\item `Convexity': $\qrom{\sum_k \lambda_k \rho_{t}^{(k)}}{\sigma_r} \le \sum_{k} \lvert \lambda_k \rvert \qrom{\rho_{t}^{(k)}}{\sigma_r}$.
\item RoM recovery: let $\sigma_r$ be a mixture of stabilizer states, i.e., $\sigma_r = \sum_k p_k s_k$ with $p_k \ge 0$ and $s_k$ $r$-qubit stabilizer states. Then $\qrom{\rho_t}{\sigma_r} = \mathcal{R}(\rho_t)$.
\end{enumerate}
\begin{proof} 
${}$ \newline
P1. It follows immediately from the fact that  $\mathcal{Q}_t(\sigma_r)$ is a convex set and that $\mathrm{Tr}(\rho_t)=\mathrm{Tr}(\eta_x)= 1$.  
\newline
P2. The optimal decompositions associated with $\qrom{\rho_{t_1}}{\sigma_{r_1}}$ and $\qrom{\rho_{t_2}}{\sigma_{r_2}}$ induce a block-decomposition $\rho_{t_1} \otimes \rho_{t_2} = \sum_x q_x \eta_x$ with each $\eta_x$ a product of a state in $\mathcal{Q}_{t_1}(\sigma_{r_1})$ and one in $\mathcal{Q}_{t_2}(\sigma_{r_2})$. The negativity of this decomposition is $\lVert q \rVert_1 = \qrom{\rho_{t_1}}{\sigma_{r_1}} \qrom{\rho_{t_2}}{\sigma_{r_2}}$. The property follows by noticing that, for every $x$, \mbox{$\eta_x \in  \mathcal{Q}_{t_1+t_2}(\sigma_{r_1} \otimes \sigma_{r_2})$}.
\newline
P3. Let $\rho_t= \sum_x q_x \eta_x$ an optimal decomposition, which satisfies $\sum_x |q_x| = \qrom{\rho_t}{\sigma_t}$ and $\eta_x \in \mathcal{Q}_t(\sigma_r)$ for every $x$. Then $\mathcal{E}(\rho_t) = \sum_x q_x \mathcal{E}(\eta_x)$ is an allowed (albeit not necessarily optimal) decomposition of $\mathcal{E}(\rho_t)$ using states $\mathcal{E}(\eta_x) \in \mathcal{Q}_t(\mathcal{E}(\sigma_r))$. Then $\qrom{\mathcal{E}(\rho_t)}{\mathcal{E}(\sigma_t)} \leq \qrom{\rho_t}{\sigma_t}$. 
Furthermore, $\mathcal{Q}_t(\mathcal{E}(\sigma_r)) \subseteq \mathcal{Q}_t(\sigma_r)$, so $\qrom{\rho_t}{\mathcal{E}(\sigma_r)} \ge \qrom{\rho_t}{\sigma_r}$.
\newline
P4 The first equality follows immediately from P3. In fact, P3 implies $\qrom{\mathcal{E}(\rho_t)}{\sigma_r} \leq \qrom{\mathcal{E}(\rho_t)}{\mathcal{E}(\sigma_r)} \leq \qrom{\rho_t}{\sigma_r}$ for every stabilizer operation $\mathcal{E}$. Since for any Clifford unitary $\mathcal{U}$, both $\mathcal{U}$ and $\mathcal{U}^\dag$ are stabilizer operations, we have
\begin{equation}
    \mathcal{R}(\rho_t|\sigma_r) \geq \mathcal{R}(\mathcal{U}(\rho_t)|\sigma_r) \geq \mathcal{R}(\mathcal{U}^\dag \circ \mathcal{U}(\rho_t)|\sigma_r) = \mathcal{R}(\rho_t|\sigma_r).
\end{equation}
The second equality in P4 follows by noticing that Clifford operations leave the set $\mathcal{Q}_t(\sigma_r)$ unchanged, i.e., $\mathcal{Q}_t(\sigma_r) = \mathcal{Q}_t(\mathcal{U}(\sigma_r))$.
\newline
P5. Let $\rho_{t}^{(k)} = \sum_x q_{x}^{(k)} \eta_{x}$ be the optimal decompositions associated with the QRoMs $\qrom{\rho_{t}^{(k)}}{\sigma_r} = \sum_{x} \lvert q_{x}^{(k)} \rvert$. They induce a decomposition of $\sum_k \lambda_k \rho_{t}^{(k)}$ as 
\begin{equation}
\sum_k \lambda_k \rho_{t}^{(k)} = \sum_x w_{x} \eta_{x}, 
\end{equation}
with $w_x = \sum_k \lambda_k q_{x}^{(k)}$. This implies that 
\begin{multline}
\mathcal{R} \biggl(\sum_k \lambda_k \rho_{t}^{(k)} \bigl \lvert \sigma_r \biggr) \le \sum_{x} \lvert w_x \rvert \le \\ \sum_{k} \lvert \lambda_k \rvert \sum_{x} \lvert q_x^{(k)} \rvert  = \sum_{k} \lvert \lambda_k \rvert \qrom{\rho_{t}^{(k)}}{\sigma_r}.
\end{multline}
\newline 
P6. Since $\sigma_r$ is a stabilizer state, by definition $\mathcal{Q}_t(\sigma_r)$ coincides with the set of all stabilizer states. Then the definition of the QRoM reduces to that of the RoM.
\end{proof}

\section{Overview of some basic notions}
\label{app:reminder}

It is well-known that the Clifford group is generated by the single-qubit Hadamard gate $H = (X+Z)/\sqrt{2} $, the phase gate $S = Z^{1/2} = \ket{0}\bra{0} + i \ket{1}\bra{1}$ acting on all possible qubits and by the CNOT gate, $\mathrm{CNOT} = \ket{0}\bra{0} \otimes I + \ket{1}\bra{1} \otimes X$ acting on all possible pairs of qubits~\cite{AaronsonGottesman}. The addition of the $T$ gate
\begin{equation}
T = \begin{pmatrix}
1 & 0 \\
0 & e^{i \pi/4}
\end{pmatrix}
\end{equation}
acting on any qubit promotes the set $\{H, S, \mathrm{CNOT}\}$ to a universal gate set \cite{nielsenChuang}. 

\begin{figure}[h]
    \centering
    \includegraphics[width=0.4 \textwidth]{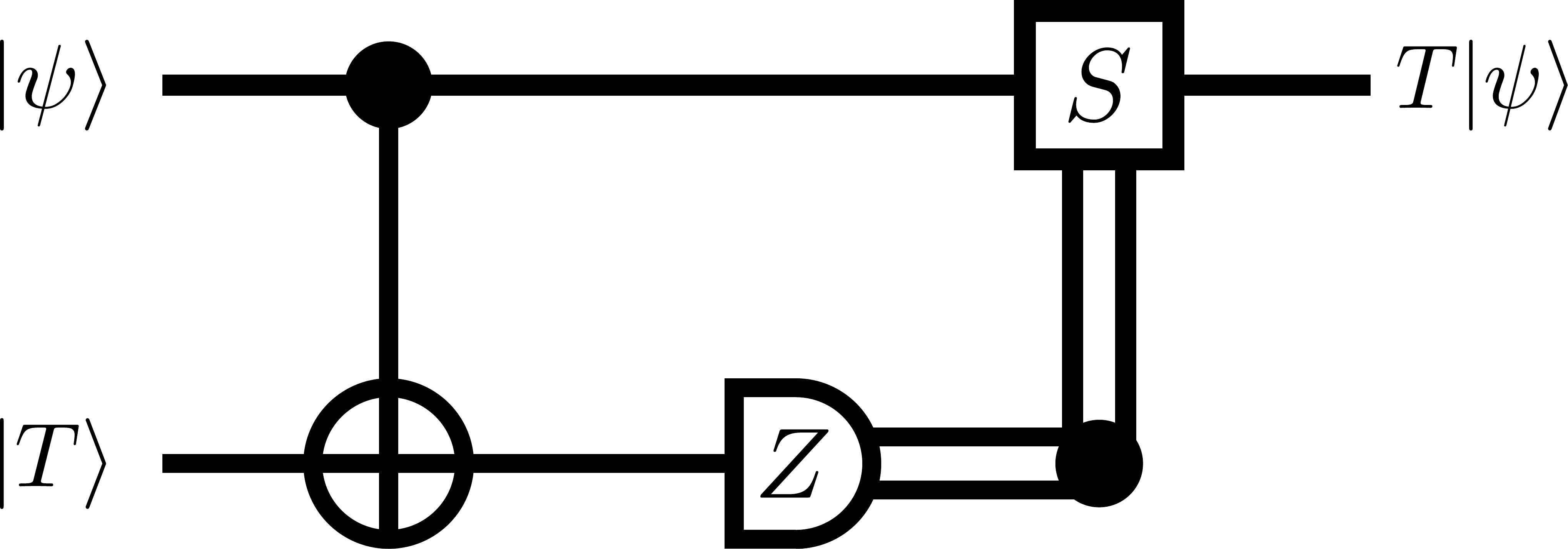}
    \caption{Gate teleportation circuit with data qubit in a generic state $\ket{\psi}$ and auxiliary state in the magic state $\ket{\magic}$. The output of the $Z$ measurement on the auxiliary qubit can  be $\lambda_Z = \pm 1$. If $\lambda_Z =1$ the identity is applied to the data qubit, while if $\lambda_Z = -1$ the phase gate $S$ is applied. This protocol applies a $T$ gate to the data qubit.}
    \label{fig:state_inj}
\end{figure}

The $T$-gate can also be obtained by injecting the magic state 
\begin{equation}
\ket{\magic} = T \ket{+}= \frac{1}{\sqrt{2}} \bigl(\ket{0} + e^{i \pi/4} \ket{1} \bigl).
\end{equation}
into the adaptive Clifford circuit shown in Fig. \ref{fig:state_inj}, a procedure known as ``gadgetization'' \cite{leung2000, bravyiKitaev, campbell2017roads}.

\section{Overhead and QRoM}
\label{app:overheadandQRoM}

Here we discuss in generality the protocol that allows to simulate the ideal resources $\rho_t$ from the available ones $\sigma_r$, with overhead given by $\qrom{\rho_t}{\sigma_r}$. Error mitigation and quantum-assisted simulation are special instances of this protocol.

We start from a decomposition of the form
\begin{equation}
    \rho_t = \sum_{x} q_{x} \eta_x, \eta_x \in \mathcal{Q}_t(\sigma_r).
\end{equation}
We then follow step 1-4 of the quasiprobability sampling protocol given in the main text. Let $P = \Pi_1 - \Pi_{-1}$. The output of the protocol is a random variable $O$ whose average $\langle O \rangle$ reads
\small
\begin{equation*}
     \sum_x \frac{|q_x|}{\|q\|_1} \sum_{p=\pm 1} \tr{\Pi_p \mathcal{A}\left(\ketbra{0}{0}^{\otimes n} \otimes \eta_x\right)} p \|q\|_1 \textrm{sign}(q_x) 
\end{equation*}
\normalsize
where $\eta_x$ is a $t$-qubit state and $\mathcal{A}$ is the (adaptive) Clifford circuit to be simulated, including the state injection gadgets. It is not difficult to see that the previous expression equals
\begin{equation}
    \tr{P \mathcal{A} \left(\ketbra{0}{0}^{\otimes n} \otimes \rho_t\right)} = \langle P \rangle. 
\end{equation}
Hence $O$ is an unbiased estimator of $\langle P \rangle$. 

Let us consider the case in which we repeat the protocol $M$ times. The output of each run is a random variable $O_i$, $i=1, \dots, M$ as discussed above. The random variables $O_i$ are assumed to be statistically independent. Moreover, the $O_i$ are identically distributed and $O_i \in [-\| q \|_1, + \| q \|_1]$. Thus, the range of $O_i$ is $r = 2 \| q \|_1$. Hoeffding's inequality states that \cite{hoeffding1963}

\begin{equation}
\mathrm{Prob} \biggl(\biggl \lvert \frac{1}{M} \sum_{i =1}^M (O_i - \langle O \rangle) \biggr \rvert  \ge \Delta \biggr ) \le 2 \exp \biggl(- \frac{2 M \Delta^2}{r^2} \biggr),
\end{equation}
with $\Delta$ the maximum error on the estimate of $\langle O \rangle$. We can ensure that the probability of having additive error greater
than $\Delta$ is less than an arbitrary $\epsilon \in (0,1]$ by requiring
that $2 \exp(-2 M \Delta^2 /r^2) \le \epsilon$, which implies $M \ge 2 \ln (2/ \epsilon) \| q \|_1^2/\Delta^2$. The QRoM is defined by the optimal decomposition achieving the minimum value of $\|q\|_1$.

\section{Lower bound on QRoM}
\label{app:lowerbound}

Here we derive a useful, general lower bound on $\qrom{\tau^{\otimes t}}{\tau_{\delta}^{\otimes r}}$ defined in Eq.  3 in the main text. Let $\rho$ be an $n$-qubit density matrix. All we need to derive a bound on the QRoM is to assume the existence of a convex magic monotone, i.e. a functional $\mathcal{M}$  with the following standard properties:
\begin{enumerate}
    \item `Convexity': $\monotone (\sum_{k}\lambda_k \rho^{(k)}) \le  \sum_x \lvert \lambda_k \rvert \monotone(\rho^{(k)})$, $\lambda_k \in \mathbb{C}$.
    \item Monotonicity: $\monotone(\mathcal{E}(\rho)) \le \monotone(\rho)$ for every stabilizer operation $\mathcal{E}$. 
    %\item Magic witness: $\monotone(\rho) \ge 1$ with equality holding if and only if $\rho$ is a stabilizer state.
\end{enumerate}
Let us consider the optimal decomposition associated with $\qrom{\tau^{\otimes t}}{\tau_{\delta}^{\otimes r}}$, which can always be written as
\begin{equation}
\tau^{\otimes t} = \sum_x q_x \mathcal{E}_x(\tau_{\delta}^{\otimes r} \otimes \ket{0} \bra{0}^{\otimes t - r}),
\end{equation}
with $\mathcal{E}_x$ stabilizer operations. Then the magic monotone $\monotone(\tau^{\otimes t})$ satisfies 
\begin{align}
\monotone(\tau^{\otimes t}) &= \monotone \biggl( \sum_x q_x \mathcal{E}_x(\tau_{\delta}^{\otimes r} \otimes \ket{0} \bra{0}^{\otimes t - r}) \biggr) \\
& \le \sum_x \lvert q_{x} \rvert \monotone (\mathcal{E}_x(\tau_{\delta}^{\otimes r} \otimes \ket{0} \bra{0}^{\otimes t - r})) \\
& \le \sum_x \lvert q_{x} \rvert \monotone (\tau_{\delta}^{\otimes r} \otimes \ket{0} \bra{0}^{\otimes t - r}) \\
& \le \monotone(\tau_{\delta}^{\otimes r}) \sum_x \lvert q_{x} \rvert=  \monotone(\tau_{\delta}^{\otimes r}) \qrom{\tau^{\otimes t}}{\tau_{\delta}^{\otimes r}}. 
\end{align}
We thus obtain the general lower bound
\begin{equation}
 \qrom{\tau^{\otimes t}}{\tau_{\delta}^{\otimes r}} \ge \frac{\monotone(\tau^{\otimes t})}{\monotone(\tau_{\delta}^{\otimes r})}.
\end{equation}

Noticing that the RoM satisfies all the assumed properties of $\mathcal{M}$ \cite{howardCampbell}, we can use it to bound the QRoM. For instance for $t=r=1$ we obtain, for $\delta \le \delta_{\mathrm{th}}$, $\mathcal{R}(\tau)/\mathcal{R}(\tau_{\delta})=1/(1 - \delta)$, which shows that the decomposition in Eq.~(7) in the main text is optimal. However, in order to obtain bounds for large $t$ using the RoM, we need to evaluate $\mathcal{R}(\tau_{\delta}^{\otimes t})$, which gets harder as $t$ increases. 

More convenient bounds can be obtained by considering magic monotones that are multiplicative at least for tensor products of single-qubit states. For instance, a monotone that has this property is the dyadic negativity introduced in Ref.~\cite{seddon2020} and there denoted by $\Lambda$. In terms of the dyadic negativity the bound on the QRoM becomes
\begin{equation}
 \qrom{\tau^{\otimes t}}{\tau_{\delta}^{\otimes r}} \ge \frac{\Lambda(\tau)^t}{\Lambda(\tau_{\delta})^r}.
\end{equation}
We obtain $\Lambda(\tau_{\delta})$ by finding an explicit dyadic decomposition that matches the numerically optimum for every $\delta$:
\begin{equation}
    \tau_\delta = q_1 \ketbra{+}{+} + q_1 \ketbra{i}{i} + q_2 \ketbra{+}{i} + q_2^* \ketbra{i}{+},
\end{equation}
where 
\begin{align}
    q_1 = & 1 -\frac{1}{\sqrt{2}} + \frac{\delta }{\sqrt{2}} \\
    q_2 = & \left(\frac{1}{2}+\frac{i}{2}\right)\left( -\sqrt{2} \delta +\sqrt{2}-1\right) 
\end{align}
 We get
\begin{equation}\label{eq:dyadnegtau}
\Lambda(\tau_{\delta}) = \begin{cases}
1, \quad &\delta \ge \delta_{\mathrm{th}}, \\
2(2 - \sqrt{2})(1-\delta/2), \quad &\delta \le \delta_{\mathrm{th}},
\end{cases}
\end{equation}
and, setting $t=r$, we obtain the bound reported in the main text.

\section{Analytical decompositions and expressions for the upper bounds on the QRoM}
\label{app:analytic}

In this section we define the pure stabilizer states:
\begin{equation}
    \ket{s_1} = \frac{1}{\sqrt{2}} (\ket{01} - \ket{10}), \quad \ket{s_2} = \frac{1}{\sqrt{2}}(\ket{00}- i \ket{11})
\end{equation}
and the mixed stabilizer states
\begin{equation}
    \eta_{\pm} = \frac{1}{2}\ketbra{\pm}{\pm} + \frac{1}{2} \ketbra{\pm i}{\pm i}.
\end{equation}

In what follows we give details of the decompositions of $\tau^{\otimes t}$ in terms of $\tau_{\delta}^{\otimes r}$ for the different $t$ and $r$ presented in the main text.

\subsection{Decomposition for $t=2, r=1$}

Consider the four-states decomposition
\begin{equation}
    \tau^{\otimes 2} = \sum_{a=1}^4 q_a \eta_a.
\end{equation}
Here $\eta_i = \ketbra{s_i}{s_i} $ for $i=1,2$
and
\begin{equation}
    \eta_3 = \frac{1}{2} \left(\tau_\delta \otimes \eta_+ + \eta_+ \otimes \tau_\delta\right),
\end{equation}
\begin{equation}
    \eta_4 = \frac{1}{2} \left(Z \tau_\delta Z \otimes \eta_- + \eta_- \otimes Z\tau_\delta Z \right).
\end{equation}

Set the coefficients in the decomposition as
\begin{align}
q_{1} = q_{2} = & \frac{1}{2} - \frac{\sqrt 2}{1 + \sqrt 2 - \delta}\\
    q_{3} = & \frac{2(1 + \sqrt 2) (-2 + \delta)}{-4 - 
 3 \sqrt 2 + (4 + 3 \sqrt 2) \delta - 2 \delta^2}, \\
 q_{4} = & \frac{-2\delta + 
  2 \sqrt 2 \delta}{-4 - 3 \sqrt 2 + 4 \delta + 
   3 \sqrt 2 \delta - 2 \delta^2}
\end{align}
The negativity of this decomposition is
\begin{equation}
    \|q \|_1 = 
    \frac{4 + 5 \sqrt 2 - \sqrt 2 \delta - 2 \delta^2}{(4 + 3 \sqrt 2)(1 -
  \delta + 2 \delta^2)}.
\end{equation}  

This coincides with the result obtained from a full numerical optimization over a subset of $\mathcal{Q}_2(\tau_{\delta}^{\otimes 1})$, which includes all two-qubit stabilizer states and all two-qubit Cliffords applied to $\tau_{\delta} \otimes \ket{0}\bra{0}$.

\subsection{Decomposition for $t=2, r=2$}
Consider the four-states decomposition
\begin{equation}
    \tau^{\otimes 2} = \sum_{a=1}^4 q_a \eta_a.
\end{equation}
Here  $\eta_i = \ketbra{s_i}{s_i}$ for $i=1,2$ and
\begin{equation}
    \eta_3 = \tau_\delta \otimes \tau_\delta, \quad \eta_4 = Z \tau_\delta Z \otimes Z \tau_\delta Z.
\end{equation}
Set the coefficients in the decomposition as
\begin{align}
q_{1} = q_{2} = & \frac{(-2 + \delta) \delta}{2 (2 - 2 \delta + \delta^2)}\\
    q_{3} = & \frac{(2-\delta)^2}{2 (2 - 4 \delta +
   3 \delta^2 - \delta^3)}, \\
 q_{4} = & - \frac{\delta^2}{2 (2 - 4 \delta + 3 \delta^2 - \delta^3)}.
\end{align}

The negativity of this decomposition is
\begin{equation}
    \|q \|_1 = = \frac{2 - 2 \delta^2 + \delta^3}{2 - 4 \delta + 
 3 \delta^2 - \delta^3} :=\overline{\mathcal{R}}(\tau^{\otimes 2} \lvert \tau_{\delta}^{\otimes 2}).
\end{equation}  
This coincides with the result obtained from a full numerical optimization over a subset of $\mathcal{Q}_2(\tau_{\delta}^{\otimes 2})$, which includes all two-qubit stabilizer states and all two-qubit Cliffords applied to $\tau_{\delta} \otimes \ket{0}\bra{0}$ and $\tau_{\delta}^{\otimes 2}$.

\subsection{Decomposition for $t=3$, $r=1$} \label{subsec:t3r1}

In this instance we perform a numerical optimization involving three-qubit stabilizer states, as well as all tensor products of single-qubit Cliffords applied to $\tau_{\delta} \otimes \ket{\Phi_s}\bra{\Phi_s}$ (and its swapped versions) with $\ket{\Phi_s}$ a two-qubit stabilizer state. We find a decomposition whose negativity is 
\begin{equation}
    \frac{7.025 + 44.899 \delta - 11.055 \delta^2 - 6.682 \delta^3}{4.025 + 22.561 \delta - 
   25.664 \delta^2 + 6.682 \delta^3},
\end{equation}
which is then an upper bound for $\qrom{\tau^{\otimes 3}}{\tau_\delta}$. The analytic expression results from a fit of the numerics and the approximation is within $10^{-4}$.

\subsection{Decomposition for $t=3, r=3$}

Let $\xi_{12} = \frac{1}{2} \ketbra{s_1}{s_1}_{12} + \frac{1}{2} \ketbra{s_2}{s_2}_{12}$, where the label `12' refers to the fact that $\xi_{12}$ has support on qubits $1$ and $2$. Similarly we define $\xi_{23}$ and $\xi_{12}$.

Consider the four-state decomposition
\begin{equation}
    \tau^{\otimes 3} = \sum_{a=1}^4 q_a \eta_a.
\end{equation}

\begin{equation}
    \eta_1 = \frac{1}{3}(\tau_{\delta,1} \otimes \xi_{23} +  \xi_{13} \otimes \tau_{\delta,2} + \xi_{12} \otimes \tau_{\delta,3}),  
\end{equation}
\begin{equation}
    \eta_2 = \frac{1}{3}(Z\tau_{\delta,1}Z \otimes \xi_{23} +  \xi_{13} \otimes Z\tau_{\delta,2} Z + \xi_{12} \otimes Z\tau_{\delta,3}Z),
\end{equation}
\begin{equation}
    \eta_3 = \tau_{\delta} \otimes \tau_\delta \otimes \tau_\delta,
\end{equation}
\begin{equation}
    \eta_4 = Z\tau_{\delta}Z \otimes Z\tau_\delta Z \otimes Z\tau_\delta Z,
\end{equation}

Set the coefficients in the decomposition as
\begin{align}
q_{1} = & \frac{3 (\delta -2)^4 \delta }{2 (\delta -1) ((\delta -2) \delta +4) (3 (\delta -2) \delta +4)},
\\
    q_{2} = & \frac{3 (\delta -2) \delta ^4}{2 (\delta -1) ((\delta -2) \delta +4) (3 (\delta -2) \delta +4)},
    \\
 q_{3} = & \frac{2 (\delta -2)^3}{(\delta -1) ((\delta -2) \delta +4) (3 (\delta -2) \delta +4)},
 \\
 q_{4} = & \frac{2 \delta ^3}{(\delta -1) ((\delta -2) \delta +4) (3 (\delta -2) \delta +4)}.
\end{align}

The negativity of this decomposition is
\begin{equation}
    \|q \|_1 = \frac{4 +6 \delta - 3 \delta ^2}{4 - 6 \delta + 3 \delta^2 - \delta^3} := \overline{\mathcal{R}}(\tau^{\otimes 3}| \tau_\delta^{\otimes 3}).
\end{equation}  

The decomposition is optimal when considering the subset defined in Subsec. \ref{subsec:t3r1} augmented with the states obtained by applying all tensor products of single-qubit Cliffords to $\tau_{\delta}^{\otimes 2} \otimes \ket{0}\bra{0}$ (and its swapped versions) and to $\tau_{\delta}^{\otimes 3}$.

\section{Error mitigation with noisy Cliffords}
\label{app:noisycliffords}

We now study a scenario where a noisy $T$-state $\tau_\delta$ enters the state injection circuit in Fig.~\ref{fig:state_inj}, but the Clifford unitaries involved are affected by depolarising noise. Specifically we assume that all the Cliffords on the data qubits, as well as each of the (two) gates involved in the state injection circuit in Fig.~\ref{fig:state_inj}, have  probability $1-\delta_c$ of working correctly, and probability $\delta_c$ of fully depolarizing the state on which they are acting. 

Let's first focus on the noisy state injection circuit. Instead of performing an ideal $T$-gate $\mathcal{T}(\cdot) = T (\cdot) T^\dag$, upon injecting a noisy $T$-state $\mathcal{\tau_\delta}$ we realize the completely-positive trace-preserving map
\begin{equation}
\nonumber
    \mathcal{T}_{\delta,\delta_c} = (1-\delta)(1-\delta_c)^2 \mathcal{T} + \delta (1-\delta_c)^2 \mathcal{D} + [1-(1-\delta_c)^2] \mathcal{G}, \end{equation}
where $\mathcal{D}$ is the dephasing channel (a map removing the off-diagonal elements in the computational basis) and $\mathcal{G}$ is the depolarising channel (transforming every state into the maximally mixed state $I/2$). Note that the channel depends on $\delta_c$, since the injection circuit itself is faulty.

To error mitigate the above channel, consider the quasiprobability decomposition
\begin{equation}
    \mathcal{T}=q_1 \mathcal{T}_{\delta,\delta_c}+q_2 \mathcal{Z} \circ \mathcal{T}_{\delta, \delta_c}+q_3 \mathcal{G},
\end{equation}
where 
$q_1=\frac{(2-\delta)}{2 (1-\delta) (1-\delta_c)^2}$,  $q_2= \frac{\delta}{2 (\delta-1) (1-\delta_c)^2}$, $q_3= \frac{\delta_c(\delta_c -2)}{(1-\delta_c)^2}$.
Note that all three operations appearing in the decompositions can be performed under our noise model. In particular, $\mathcal{Z} \circ \mathcal{T}_{\delta, \delta_c}$ can be realized by injecting a rotated noisy $T$-state $(1-\delta) Z \tau_\delta Z + \delta I/2$ into faulty version of the circuit in Fig.~\ref{fig:state_inj}. 

Whenever a $T$-gate $\mathcal{T}$ appears in the circuit, we instead perform $\mathcal{T}_{\delta, \delta_c}$ with probability $q_1/\|q\|_1$, $\mathcal{Z} \circ \mathcal{T}_{\delta, \delta_c}$ with probability $q_2/\|q\|_1$ and $\mathcal{G}$ with probability $q_3/\|q\|_1$ (the latter can be realized by random single qubit Paulis). The error mitigation overhead (compared to the ideal circuit) is quantified by $ \|q\|^{2t}_1$, with $t$ the $T$-count, where
\begin{equation}
    \|q\|_1= \frac{2-\delta}{(1-\delta) (1-\delta_c)^2}-1.
\end{equation}
Note that for $\delta_c = 0$ we recover $\|q\|_1 = 1/(1-\delta)= \mathcal{R}(\tau|\tau_\delta)$, as expected from our analysis of error mitigation of the noisy $T$-state with ideal Cliffords. 

Now let us consider the $k=1,2$ qubit Clifford unitaries $\mathcal{U}^{(k)}$ on the data qubits. These can be mitigated using the quasiprobability decomposition discussed in Refs.~\cite{temmeBravyi,takagi2020}:
\begin{equation}
    \mathcal{U}^{(k)} = s_1 \mathcal{U}^{(k)}_{\delta_c} + s_2 \left(\frac{1}{2^{2k} -1} \sum_{i=1}^{2^{2k}-1} \mathcal{P}^i_{\delta_c}\right),  
\end{equation}
where $\mathcal{U}^{(k)}_{\delta_c} = (1-\delta_c) \mathcal{U}^{(k)} + \delta_c  \mathcal{G}_k$ ($\mathcal{G}_k(\rho) = I/2^k$ for all $\rho$), and $\mathcal{P}^i_{\delta_c} = (1-\delta_c) \mathcal{P}^i + \delta_c  \mathcal{G}_k$, where $\mathcal{P}^i$ are all single or two-qubit Pauli unitaries (excluding the identity). The coefficients are 
\begin{equation}
    s_1 = 1+ \frac{(2^{2k} -1)\delta_c}{2^{2k}(1-\delta_c)}, \quad  s_2 = - \frac{(2^{2k} -1) \delta_c}{2^{2k} (1-\delta_c)}.
\end{equation}   

Whenever a single qubit Clifford unitary is performed on the data qubits, we instead perform a noisy version $\mathcal{U}_{\delta_c}$ with probability $s_1/\|s\|_1$ and, with probability $s_2/\|s\|_1$, we perform a random noisy Pauli $\mathcal{P}^i_{\delta_c}$. The error mitigation overhead (compared to the ideal circuit) is quantified by $ \|s\|^{2n^{(k)}_c}_1$, with $n^{(k)}_c$ the number of $k=1,2$ qubit Cliffords. In particular
\begin{equation}
    \|s\|_1= \frac{1+\delta_c/2}{1-\delta_c}, \quad \textrm{for } k=1,
\end{equation}
\begin{equation}
    \|s\|_1= \frac{1+7\delta_c/8}{1-\delta_c}, \textrm{for } k=2.
\end{equation}
The total overhead for error mitigation of a circuit involving $n^{(1)}_c$ single qubit Cliffords and $n^{(2)}_c$ two-qubit Cliffords on the data qubits, as well as $t$ $T$-gates, is
\small
\begin{equation}
\nonumber
  \left( \frac{2-\delta}{(1-\delta) (1-\delta_c)^2}-1 \right)^{2t} \left(\frac{1+\delta_c/2}{1-\delta_c}\right)^{2 n^{(1)}_c} \left(\frac{1+7\delta_c/8}{1-\delta_c}\right)^{2 n^{(2)}_c}.
\end{equation}

\section{Mitigating errors in magic state distillation protocols}
\label{app:mitigate_dist}
In the main text, we analyzed how magic state distillation can increase the number of Cliffords and $T$-gates that can be mitigated with a given overhead. In particular, Fig. 3 (left) was obtained taking as input the residual $T$-state noise after $m$ distillation rounds $(m=1,2)$ of the Bravyi-Haah $14 \mapsto 2$ protocol \cite{bravyiHaah}. Implicitly, this assumes the standard analysis of the distillation stage, where Cliffords are taking to be noiseless. Here, we show how this assumption can be lifted. Specifically, we apply the same quasiprobability-based protocol discussed in the main text to mitigate the error on the Cliffords required by the distillation protocol. 

As in the main text, we focus on the Bravyi-Haah protocols based on triorthogonal matrices. We refer the reader to the original paper in Ref.~\cite{bravyiHaah} for all the technical details. A general $n \mapsto k$ Bravyi-Haah protocol requires $n$ input noisy $T$-states and $n$ input target qubits starting in the reference state $\ket{+}^{\otimes k} \otimes \ket{0}^{\otimes (n - k)}$. The essential operations involved in the protocol are the following:
\begin{enumerate}
    \item Encoding of the target qubits in $k$ logical `$+$ states' of a suitable CSS code \cite{nielsenChuang}.
    \item Application of $n$ noisy $T$-gates, using the $n$ noisy $T$-states in the gate teleportation gadget of Fig. \ref{fig:state_inj}.
    \item Application of a Clifford gate composed only of $\mathrm{CZ} = \ketbra{0}{0} \otimes I + \ketbra{1}{1} \otimes Z$, $S$ and $S^\dag$ elementary gates. 
    \item $X$ syndrome measurements associated with the CSS code with post-selection on the output.
    \item Decoding of the CSS code.
\end{enumerate}
Encoding and decoding for general stabilizer codes, and thus also for CSS codes, can be implemented with Clifford gates. Let $r$ be the rank (over $\mathrm{GF(2)})$ of the $X$ part of the $k \times 2n$ stabilizer generator matrix of the encoding unitary (the rank of the first $n$ columns).  It is known that the number of two-qubit Cliffords needed to encode in a stabilizer code is upper bounded by $k(n-k-r)+r(n-1)$, and the number of one-qubit ones is bounded by $r$ (see Chap. 4 in Ref. \cite{gottesmanThesis}). Since decoding is the inverse operation of encoding, the number of single and two-qubit Cliffords for these operations is at most $2k(n-k-r) + 2rn$. For the $14 \mapsto 2$ Bravyi-Haah protocol (which has $r=3$), assuming that we can directly perform $\mathrm{CZ}, S$ and $S^{\dagger}$ only, $10$ elementary gates are needed in step 3 of the protocol and at most $120$ for encoding and decoding. Thus, we obtain that the total number of elementary Cliffords is upper bounded by $n_{c, d}^{\mathrm{max}} =  130$. 

We assume that also these additional Cliffords can be affected by depolarizing noise as in Eq.(9) in the main text, with noise parameter $\delta_{c, d}$ (where $d$ stands for `distillation'). In what follows, we assume that also the Cliffords needed in the distillation protocol are error mitigated using the same gate mitigation protocol (1, 2', 3, 4) discussed in the main text. This cancels the noise in the elementary unitaries used for the distillation stage, at the price of an additional sampling overhead. As in the main text, we assume the worst-case scenario in which such unitaries are all treated as two-qubit gates. For this protocol, we obtain that the upper bound on the sampling overhead at the $m$-th distillation round is

\begin{figure}[h!]
   \includegraphics[width=0.378 \textwidth]{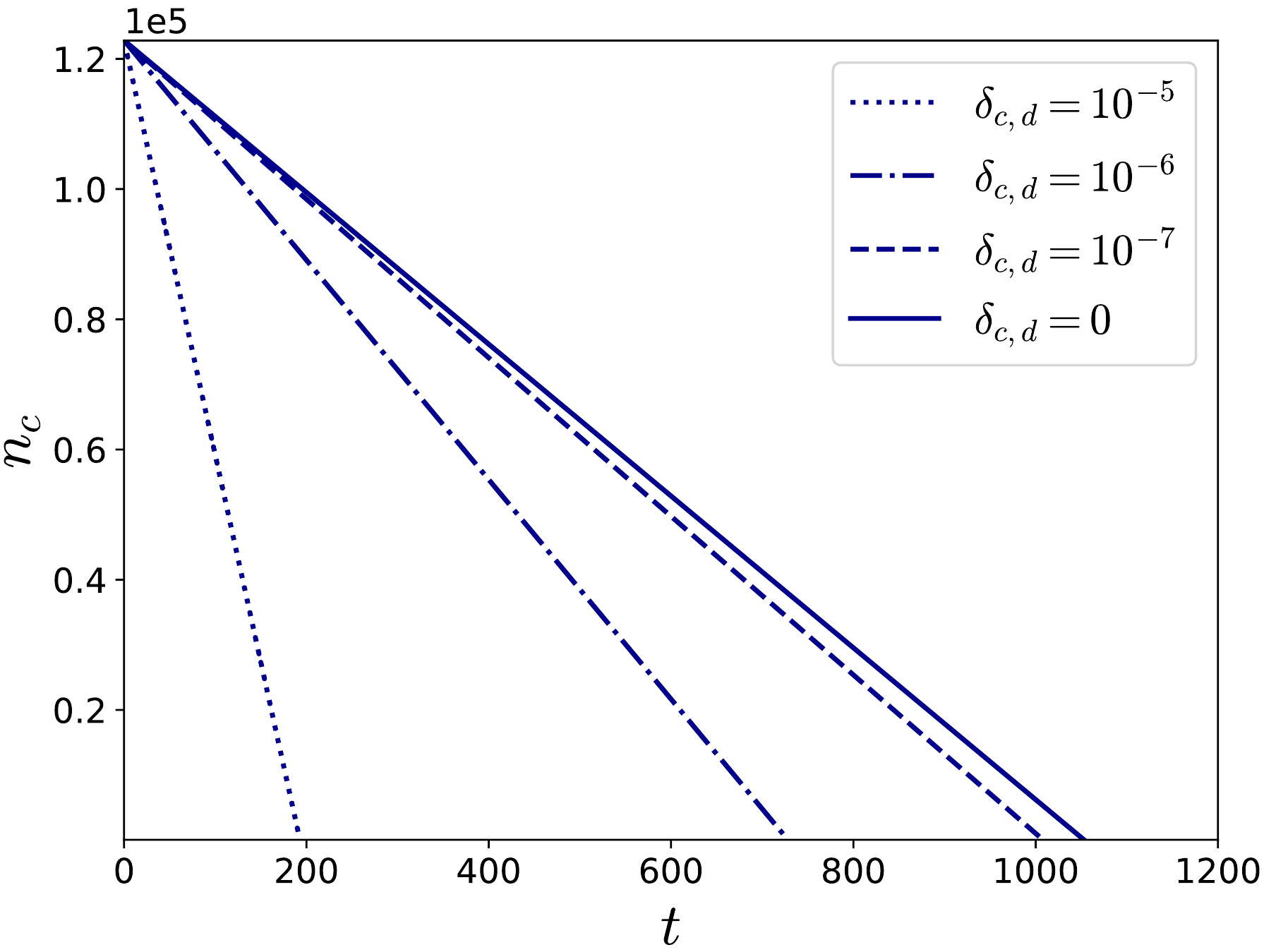}
    \caption{\emph{Error mitigation and noisy distillation}. Maximum number of $T$-gates ($t$) and one or two-qubit Cliffords ($n_c$) that can be error mitigated with moderate overhead $(\leq 100)$ after two rounds rounds of Bravyi-Haah $14\mapsto 2$ distillation.  The residual $T$-state depolarizing noise after two rounds of distillation is taken to be $\delta \approx 2 \times 10^{-3}$, and the Clifford unitaries on the data qubits are assumed to be affected by depolarizing noise $\delta_c = 10^{-5}$, as in the main text. One and two-qubit gates of the distillation stage are taken to be affected by depolarizing noise $\delta_{c,d} \in \{ 0, 10^{-7}, 10^{-6}, 10^{-5}$ \}.  All noise sources (including those of the distillation stage) are error mitigated with the  protocol (1, 2', 3, 4) presented in Section `Error  mitigation  with  noisy  Cliffords' of the main text. } 
    \label{fig:noisy_distillation}
\end{figure}

\small
\begin{multline}
\nonumber
   M \leq \left(\frac{2-\delta_m}{(1-\delta_m) (1-\delta_c)^2}-1\right)^{2 t}\left(\frac{1+\frac{7 \delta_c}{8}}{1-\delta_c}\right)^{2 n_c} \times  \\
   \left(\frac{1+\frac{7 \delta_{c, d}}{8}}{1-\delta_{c, d}}\right)^{2 n_{c, d}^{\mathrm{max}} \sum_{\ell =1}^m t^{(l)}},
\end{multline}
where $\delta_m = 7 \delta_{m-1}^2 $ is the error after the $m$-th distillation round, $\delta_0 = \delta$ is the noise on the $T$-states before distillation, and we defined $t^{(l)} = 14 t^{(l-1 )}/2$, with $t^{(0)} = t$ the number of $T$ gates. Two rounds of distillation are sufficient to enter classically intractable regimes when one adds to $\delta_c=10^{-5}$ an error $\delta_{c,d}=10^{-5}$ (see Fig.~\ref{fig:noisy_distillation}). Due to the large number of elementary gates involved in the distillation stage, noise is of course rather detrimental to the overall performance.

Our analysis illustrates the potential idea of applying error mitigation techniques to the distillation protocol itself. However, a much more detailed analysis is required to investigate the potential impact of this idea. On the one hand, our current analysis is based on a general, and potentially loose, upper bound on the number of elementary gates required in the encoding and decoding stages of the distillation protocol. Then, one can anticipate lower overheads in the context of specific, optimized implementations of distillation protocols.
On the other hand, a more realistic analysis would require lifting further implicit idealizations (e.g, ideal measurements, noise beyond depolarisation model) which will increase the overhead. Hence, we leave to future work a more detailed analysis of error mitigation in distillation protocols in the context of specific implementations.

\bibliography{biblio_errormitigation.bib}

\begin{thebibliography}{46}%
\makeatletter
\providecommand \@ifxundefined [1]{%
 \@ifx{#1\undefined}
}%
\providecommand \@ifnum [1]{%
 \ifnum #1\expandafter \@firstoftwo
 \else \expandafter \@secondoftwo
 \fi
}%
\providecommand \@ifx [1]{%
 \ifx #1\expandafter \@firstoftwo
 \else \expandafter \@secondoftwo
 \fi
}%
\providecommand \natexlab [1]{#1}%
\providecommand \enquote  [1]{``#1''}%
\providecommand \bibnamefont  [1]{#1}%
\providecommand \bibfnamefont [1]{#1}%
\providecommand \citenamefont [1]{#1}%
\providecommand \href@noop [0]{\@secondoftwo}%
\providecommand \href [0]{\begingroup \@sanitize@url \@href}%
\providecommand \@href[1]{\@@startlink{#1}\@@href}%
\providecommand \@@href[1]{\endgroup#1\@@endlink}%
\providecommand \@sanitize@url [0]{\catcode `\\12\catcode `\$12\catcode
  `\&12\catcode `\#12\catcode `\^12\catcode `\_12\catcode `\%12\relax}%
\providecommand \@@startlink[1]{}%
\providecommand \@@endlink[0]{}%
\providecommand \url  [0]{\begingroup\@sanitize@url \@url }%
\providecommand \@url [1]{\endgroup\@href {#1}{\urlprefix }}%
\providecommand \urlprefix  [0]{URL }%
\providecommand \Eprint [0]{\href }%
\providecommand \doibase [0]{https://doi.org/}%
\providecommand \selectlanguage [0]{\@gobble}%
\providecommand \bibinfo  [0]{\@secondoftwo}%
\providecommand \bibfield  [0]{\@secondoftwo}%
\providecommand \translation [1]{[#1]}%
\providecommand \BibitemOpen [0]{}%
\providecommand \bibitemStop [0]{}%
\providecommand \bibitemNoStop [0]{.\EOS\space}%
\providecommand \EOS [0]{\spacefactor3000\relax}%
\providecommand \BibitemShut  [1]{\csname bibitem#1\endcsname}%
\let\auto@bib@innerbib\@empty
%</preamble>
\bibitem [{\citenamefont {Campbell}\ \emph {et~al.}(2017)\citenamefont
  {Campbell}, \citenamefont {Terhal},\ and\ \citenamefont
  {Vuillot}}]{campbell2017roads}%
  \BibitemOpen
  \bibfield  {author} {\bibinfo {author} {\bibfnamefont {E.~T.}\ \bibnamefont
  {Campbell}}, \bibinfo {author} {\bibfnamefont {B.~M.}\ \bibnamefont
  {Terhal}},\ and\ \bibinfo {author} {\bibfnamefont {C.}~\bibnamefont
  {Vuillot}},\ }\bibfield  {title} {\bibinfo {title} {Roads towards
  fault-tolerant universal quantum computation},\ }\href
  {https://doi.org/10.1038/nature23460} {\bibfield  {journal} {\bibinfo
  {journal} {Nature}\ }\textbf {\bibinfo {volume} {549}},\ \bibinfo {pages}
  {172} (\bibinfo {year} {2017})}\BibitemShut {NoStop}%
\bibitem [{\citenamefont {Eastin}\ and\ \citenamefont
  {Knill}(2009)}]{eastinknill}%
  \BibitemOpen
  \bibfield  {author} {\bibinfo {author} {\bibfnamefont {B.}~\bibnamefont
  {Eastin}}\ and\ \bibinfo {author} {\bibfnamefont {E.}~\bibnamefont {Knill}},\
  }\bibfield  {title} {\bibinfo {title} {Restrictions on transversal encoded
  quantum gate sets},\ }\href {https://doi.org/10.1103/PhysRevLett.102.110502}
  {\bibfield  {journal} {\bibinfo  {journal} {Phys. Rev. Lett.}\ }\textbf
  {\bibinfo {volume} {102}},\ \bibinfo {pages} {110502} (\bibinfo {year}
  {2009})}\BibitemShut {NoStop}%
\bibitem [{\citenamefont {Aaronson}\ and\ \citenamefont
  {Gottesman}(2004)}]{AaronsonGottesman}%
  \BibitemOpen
  \bibfield  {author} {\bibinfo {author} {\bibfnamefont {S.}~\bibnamefont
  {Aaronson}}\ and\ \bibinfo {author} {\bibfnamefont {D.}~\bibnamefont
  {Gottesman}},\ }\bibfield  {title} {\bibinfo {title} {Improved simulation of
  stabilizer circuits},\ }\href {https://doi.org/10.1103/PhysRevA.70.052328}
  {\bibfield  {journal} {\bibinfo  {journal} {Phys. Rev. A}\ }\textbf {\bibinfo
  {volume} {70}},\ \bibinfo {pages} {052328} (\bibinfo {year}
  {2004})}\BibitemShut {NoStop}%
\bibitem [{\citenamefont {Seddon}\ and\ \citenamefont
  {Campbell}(2019)}]{seddonCampbell}%
  \BibitemOpen
  \bibfield  {author} {\bibinfo {author} {\bibfnamefont {J.~R.}\ \bibnamefont
  {Seddon}}\ and\ \bibinfo {author} {\bibfnamefont {E.~T.}\ \bibnamefont
  {Campbell}},\ }\bibfield  {title} {\bibinfo {title} {Quantifying magic for
  multi-qubit operations},\ }\href {https://doi.org/10.1098/rspa.2019.0251}
  {\bibfield  {journal} {\bibinfo  {journal} {Proceedings of the Royal Society
  A: Mathematical, Physical and Engineering Sciences}\ }\textbf {\bibinfo
  {volume} {475}},\ \bibinfo {pages} {20190251} (\bibinfo {year}
  {2019})}\BibitemShut {NoStop}%
\bibitem [{\citenamefont {Gottesman}(1998)}]{Gottesman:1998hu}%
  \BibitemOpen
  \bibfield  {author} {\bibinfo {author} {\bibfnamefont {D.}~\bibnamefont
  {Gottesman}},\ }\bibfield  {title} {\bibinfo {title} {{The Heisenberg
  representation of quantum computers}},\ }in\ \href@noop {} {\emph {\bibinfo
  {booktitle} {{22nd International Colloquium on Group Theoretical Methods in
  Physics}}}}\ (\bibinfo {year} {1998})\ pp.\ \bibinfo {pages} {32--43},\
  \Eprint {https://arxiv.org/abs/quant-ph/9807006} {arXiv:quant-ph/9807006}
  \BibitemShut {NoStop}%
\bibitem [{\citenamefont {Bravyi}\ and\ \citenamefont
  {Kitaev}(2005)}]{bravyiKitaev}%
  \BibitemOpen
  \bibfield  {author} {\bibinfo {author} {\bibfnamefont {S.}~\bibnamefont
  {Bravyi}}\ and\ \bibinfo {author} {\bibfnamefont {A.}~\bibnamefont
  {Kitaev}},\ }\bibfield  {title} {\bibinfo {title} {Universal quantum
  computation with ideal clifford gates and noisy ancillas},\ }\href
  {https://doi.org/10.1103/PhysRevA.71.022316} {\bibfield  {journal} {\bibinfo
  {journal} {Phys. Rev. A}\ }\textbf {\bibinfo {volume} {71}},\ \bibinfo
  {pages} {022316} (\bibinfo {year} {2005})}\BibitemShut {NoStop}%
\bibitem [{\citenamefont {Bravyi}\ and\ \citenamefont
  {Haah}(2012)}]{bravyiHaah}%
  \BibitemOpen
  \bibfield  {author} {\bibinfo {author} {\bibfnamefont {S.}~\bibnamefont
  {Bravyi}}\ and\ \bibinfo {author} {\bibfnamefont {J.}~\bibnamefont {Haah}},\
  }\bibfield  {title} {\bibinfo {title} {Magic-state distillation with low
  overhead},\ }\href {https://doi.org/10.1103/PhysRevA.86.052329} {\bibfield
  {journal} {\bibinfo  {journal} {Phys. Rev. A}\ }\textbf {\bibinfo {volume}
  {86}},\ \bibinfo {pages} {052329} (\bibinfo {year} {2012})}\BibitemShut
  {NoStop}%
\bibitem [{\citenamefont {Fowler}\ \emph {et~al.}(2013)\citenamefont {Fowler},
  \citenamefont {Devitt},\ and\ \citenamefont {Jones}}]{Fowler2013}%
  \BibitemOpen
  \bibfield  {author} {\bibinfo {author} {\bibfnamefont {A.~G.}\ \bibnamefont
  {Fowler}}, \bibinfo {author} {\bibfnamefont {S.~J.}\ \bibnamefont {Devitt}},\
  and\ \bibinfo {author} {\bibfnamefont {C.}~\bibnamefont {Jones}},\ }\bibfield
   {title} {\bibinfo {title} {Surface code implementation of block code state
  distillation},\ }\bibfield  {journal} {\bibinfo  {journal} {Scientific
  Reports}\ }\textbf {\bibinfo {volume} {3}},\ \href
  {https://doi.org/10.1038/srep01939} {10.1038/srep01939} (\bibinfo {year}
  {2013})\BibitemShut {NoStop}%
\bibitem [{\citenamefont {Meier}\ \emph {et~al.}(2013)\citenamefont {Meier},
  \citenamefont {Eastin},\ and\ \citenamefont {Knill}}]{meier2013}%
  \BibitemOpen
  \bibfield  {author} {\bibinfo {author} {\bibfnamefont {A.~M.}\ \bibnamefont
  {Meier}}, \bibinfo {author} {\bibfnamefont {B.}~\bibnamefont {Eastin}},\ and\
  \bibinfo {author} {\bibfnamefont {E.}~\bibnamefont {Knill}},\ }\bibfield
  {title} {\bibinfo {title} {Magic-state distillation with the four-qubit
  code},\ }\href@noop {} {\bibfield  {journal} {\bibinfo  {journal} {Quantum
  Info. Comput.}\ }\textbf {\bibinfo {volume} {13}},\ \bibinfo {pages}
  {195–209} (\bibinfo {year} {2013})}\BibitemShut {NoStop}%
\bibitem [{\citenamefont {Litinski}(2019)}]{litinski2019}%
  \BibitemOpen
  \bibfield  {author} {\bibinfo {author} {\bibfnamefont {D.}~\bibnamefont
  {Litinski}},\ }\bibfield  {title} {\bibinfo {title} {Magic {S}tate
  {D}istillation: {N}ot as {C}ostly as {Y}ou {T}hink},\ }\href
  {https://doi.org/10.22331/q-2019-12-02-205} {\bibfield  {journal} {\bibinfo
  {journal} {{Quantum}}\ }\textbf {\bibinfo {volume} {3}},\ \bibinfo {pages}
  {205} (\bibinfo {year} {2019})}\BibitemShut {NoStop}%
\bibitem [{\citenamefont {O'Gorman}\ and\ \citenamefont
  {Campbell}(2017)}]{ogormancampbell2017}%
  \BibitemOpen
  \bibfield  {author} {\bibinfo {author} {\bibfnamefont {J.}~\bibnamefont
  {O'Gorman}}\ and\ \bibinfo {author} {\bibfnamefont {E.~T.}\ \bibnamefont
  {Campbell}},\ }\bibfield  {title} {\bibinfo {title} {Quantum computation with
  realistic magic-state factories},\ }\href
  {https://doi.org/10.1103/PhysRevA.95.032338} {\bibfield  {journal} {\bibinfo
  {journal} {Phys. Rev. A}\ }\textbf {\bibinfo {volume} {95}},\ \bibinfo
  {pages} {032338} (\bibinfo {year} {2017})}\BibitemShut {NoStop}%
\bibitem [{\citenamefont {Temme}\ \emph {et~al.}(2017)\citenamefont {Temme},
  \citenamefont {Bravyi},\ and\ \citenamefont {Gambetta}}]{temmeBravyi}%
  \BibitemOpen
  \bibfield  {author} {\bibinfo {author} {\bibfnamefont {K.}~\bibnamefont
  {Temme}}, \bibinfo {author} {\bibfnamefont {S.}~\bibnamefont {Bravyi}},\ and\
  \bibinfo {author} {\bibfnamefont {J.~M.}\ \bibnamefont {Gambetta}},\
  }\bibfield  {title} {\bibinfo {title} {Error mitigation for short-depth
  quantum circuits},\ }\href {https://doi.org/10.1103/PhysRevLett.119.180509}
  {\bibfield  {journal} {\bibinfo  {journal} {Phys. Rev. Lett.}\ }\textbf
  {\bibinfo {volume} {119}},\ \bibinfo {pages} {180509} (\bibinfo {year}
  {2017})}\BibitemShut {NoStop}%
\bibitem [{\citenamefont {Kandala}\ \emph {et~al.}(2019)\citenamefont
  {Kandala}, \citenamefont {Temme}, \citenamefont {C{\'o}rcoles}, \citenamefont
  {Mezzacapo}, \citenamefont {Chow},\ and\ \citenamefont
  {Gambetta}}]{kandala2019error}%
  \BibitemOpen
  \bibfield  {author} {\bibinfo {author} {\bibfnamefont {A.}~\bibnamefont
  {Kandala}}, \bibinfo {author} {\bibfnamefont {K.}~\bibnamefont {Temme}},
  \bibinfo {author} {\bibfnamefont {A.~D.}\ \bibnamefont {C{\'o}rcoles}},
  \bibinfo {author} {\bibfnamefont {A.}~\bibnamefont {Mezzacapo}}, \bibinfo
  {author} {\bibfnamefont {J.~M.}\ \bibnamefont {Chow}},\ and\ \bibinfo
  {author} {\bibfnamefont {J.~M.}\ \bibnamefont {Gambetta}},\ }\bibfield
  {title} {\bibinfo {title} {Error mitigation extends the computational reach
  of a noisy quantum processor},\ }\href
  {https://doi.org/10.1038/s41586-019-1040-7} {\bibfield  {journal} {\bibinfo
  {journal} {Nature}\ }\textbf {\bibinfo {volume} {567}},\ \bibinfo {pages}
  {491} (\bibinfo {year} {2019})}\BibitemShut {NoStop}%
\bibitem [{\citenamefont {Endo}\ \emph {et~al.}(2018)\citenamefont {Endo},
  \citenamefont {Benjamin},\ and\ \citenamefont {Li}}]{endobenjamin}%
  \BibitemOpen
  \bibfield  {author} {\bibinfo {author} {\bibfnamefont {S.}~\bibnamefont
  {Endo}}, \bibinfo {author} {\bibfnamefont {S.~C.}\ \bibnamefont {Benjamin}},\
  and\ \bibinfo {author} {\bibfnamefont {Y.}~\bibnamefont {Li}},\ }\bibfield
  {title} {\bibinfo {title} {Practical quantum error mitigation for near-future
  applications},\ }\href {https://doi.org/10.1103/PhysRevX.8.031027} {\bibfield
   {journal} {\bibinfo  {journal} {Phys. Rev. X}\ }\textbf {\bibinfo {volume}
  {8}},\ \bibinfo {pages} {031027} (\bibinfo {year} {2018})}\BibitemShut
  {NoStop}%
\bibitem [{\citenamefont {Takagi}(2021)}]{takagi2020}%
  \BibitemOpen
  \bibfield  {author} {\bibinfo {author} {\bibfnamefont {R.}~\bibnamefont
  {Takagi}},\ }\bibfield  {title} {\bibinfo {title} {Optimal resource cost for
  error mitigation},\ }\href {https://doi.org/10.1103/PhysRevResearch.3.033178}
  {\bibfield  {journal} {\bibinfo  {journal} {Phys. Rev. Research}\ }\textbf
  {\bibinfo {volume} {3}},\ \bibinfo {pages} {033178} (\bibinfo {year}
  {2021})}\BibitemShut {NoStop}%
\bibitem [{\citenamefont {Endo}\ \emph {et~al.}(2021)\citenamefont {Endo},
  \citenamefont {Cai}, \citenamefont {Benjamin},\ and\ \citenamefont
  {Yuan}}]{endo2021}%
  \BibitemOpen
  \bibfield  {author} {\bibinfo {author} {\bibfnamefont {S.}~\bibnamefont
  {Endo}}, \bibinfo {author} {\bibfnamefont {Z.}~\bibnamefont {Cai}}, \bibinfo
  {author} {\bibfnamefont {S.~C.}\ \bibnamefont {Benjamin}},\ and\ \bibinfo
  {author} {\bibfnamefont {X.}~\bibnamefont {Yuan}},\ }\bibfield  {title}
  {\bibinfo {title} {Hybrid quantum-classical algorithms and quantum error
  mitigation},\ }\href {https://doi.org/10.7566/JPSJ.90.032001} {\bibfield
  {journal} {\bibinfo  {journal} {Journal of the Physical Society of Japan}\
  }\textbf {\bibinfo {volume} {90}},\ \bibinfo {pages} {032001} (\bibinfo
  {year} {2021})}\BibitemShut {NoStop}%
\bibitem [{\citenamefont {Howard}\ and\ \citenamefont
  {Campbell}(2017)}]{howardCampbell}%
  \BibitemOpen
  \bibfield  {author} {\bibinfo {author} {\bibfnamefont {M.}~\bibnamefont
  {Howard}}\ and\ \bibinfo {author} {\bibfnamefont {E.}~\bibnamefont
  {Campbell}},\ }\bibfield  {title} {\bibinfo {title} {Application of a
  resource theory for magic states to fault-tolerant quantum computing},\
  }\href {https://doi.org/10.1103/PhysRevLett.118.090501} {\bibfield  {journal}
  {\bibinfo  {journal} {Phys. Rev. Lett.}\ }\textbf {\bibinfo {volume} {118}},\
  \bibinfo {pages} {090501} (\bibinfo {year} {2017})}\BibitemShut {NoStop}%
\bibitem [{\citenamefont {Heinrich}\ and\ \citenamefont
  {Gross}(2019)}]{heinrichGross2019}%
  \BibitemOpen
  \bibfield  {author} {\bibinfo {author} {\bibfnamefont {M.}~\bibnamefont
  {Heinrich}}\ and\ \bibinfo {author} {\bibfnamefont {D.}~\bibnamefont
  {Gross}},\ }\bibfield  {title} {\bibinfo {title} {Robustness of {M}agic and
  {S}ymmetries of the {S}tabiliser {P}olytope},\ }\href
  {https://doi.org/10.22331/q-2019-04-08-132} {\bibfield  {journal} {\bibinfo
  {journal} {{Quantum}}\ }\textbf {\bibinfo {volume} {3}},\ \bibinfo {pages}
  {132} (\bibinfo {year} {2019})}\BibitemShut {NoStop}%
\bibitem [{\citenamefont {Hakkaku}\ and\ \citenamefont
  {Fujii}(2021)}]{hakkakufujii2020}%
  \BibitemOpen
  \bibfield  {author} {\bibinfo {author} {\bibfnamefont {S.}~\bibnamefont
  {Hakkaku}}\ and\ \bibinfo {author} {\bibfnamefont {K.}~\bibnamefont
  {Fujii}},\ }\bibfield  {title} {\bibinfo {title} {Comparative study of
  sampling-based simulation costs of noisy quantum circuits},\ }\href
  {https://doi.org/10.1103/PhysRevApplied.15.064027} {\bibfield  {journal}
  {\bibinfo  {journal} {Phys. Rev. Applied}\ }\textbf {\bibinfo {volume}
  {15}},\ \bibinfo {pages} {064027} (\bibinfo {year} {2021})}\BibitemShut
  {NoStop}%
\bibitem [{Note1()}]{Note1}%
  \BibitemOpen
  \bibinfo {note} {A complementary approach analyzing error mitigation of
  decoding and approximating errors in the fault-tolerant regime has been
  discussed in Ref.~\cite {suzuki2020quantum}}\BibitemShut {NoStop}%
\bibitem [{Note3()}]{Note3}%
  \BibitemOpen
  \bibinfo {note} {If $| {T} \rangle $ is affected by a different noise, it can
  be brought into this standard form by applying the identity or the Clifford
  unitary $ \protect \qopname \relax o{exp}[-i \pi /4] S X$ with probability
  $1/2$, as also remarked in Ref.~\cite {bravyiHaah}. Here $S$ is the phase
  gate and $X$ the Pauli $x$ unitary.}\BibitemShut {Stop}%
\bibitem [{\citenamefont {Takagi}\ and\ \citenamefont
  {Zhuang}(2018)}]{takagiQuntao}%
  \BibitemOpen
  \bibfield  {author} {\bibinfo {author} {\bibfnamefont {R.}~\bibnamefont
  {Takagi}}\ and\ \bibinfo {author} {\bibfnamefont {Q.}~\bibnamefont
  {Zhuang}},\ }\bibfield  {title} {\bibinfo {title} {Convex resource theory of
  non-gaussianity},\ }\href {https://doi.org/10.1103/PhysRevA.97.062337}
  {\bibfield  {journal} {\bibinfo  {journal} {Phys. Rev. A}\ }\textbf {\bibinfo
  {volume} {97}},\ \bibinfo {pages} {062337} (\bibinfo {year}
  {2018})}\BibitemShut {NoStop}%
\bibitem [{\citenamefont {Albarelli}\ \emph {et~al.}(2018)\citenamefont
  {Albarelli}, \citenamefont {Genoni}, \citenamefont {Paris},\ and\
  \citenamefont {Ferraro}}]{albarelli2019}%
  \BibitemOpen
  \bibfield  {author} {\bibinfo {author} {\bibfnamefont {F.}~\bibnamefont
  {Albarelli}}, \bibinfo {author} {\bibfnamefont {M.~G.}\ \bibnamefont
  {Genoni}}, \bibinfo {author} {\bibfnamefont {M.~G.~A.}\ \bibnamefont
  {Paris}},\ and\ \bibinfo {author} {\bibfnamefont {A.}~\bibnamefont
  {Ferraro}},\ }\bibfield  {title} {\bibinfo {title} {Resource theory of
  quantum non-gaussianity and wigner negativity},\ }\href
  {https://doi.org/10.1103/PhysRevA.98.052350} {\bibfield  {journal} {\bibinfo
  {journal} {Phys. Rev. A}\ }\textbf {\bibinfo {volume} {98}},\ \bibinfo
  {pages} {052350} (\bibinfo {year} {2018})}\BibitemShut {NoStop}%
\bibitem [{\citenamefont {Veitch}\ \emph {et~al.}(2014)\citenamefont {Veitch},
  \citenamefont {Mousavian}, \citenamefont {Gottesman},\ and\ \citenamefont
  {Emerson}}]{veitch2014resource}%
  \BibitemOpen
  \bibfield  {author} {\bibinfo {author} {\bibfnamefont {V.}~\bibnamefont
  {Veitch}}, \bibinfo {author} {\bibfnamefont {S.~A.~H.}\ \bibnamefont
  {Mousavian}}, \bibinfo {author} {\bibfnamefont {D.}~\bibnamefont
  {Gottesman}},\ and\ \bibinfo {author} {\bibfnamefont {J.}~\bibnamefont
  {Emerson}},\ }\bibfield  {title} {\bibinfo {title} {The resource theory of
  stabilizer quantum computation},\ }\href
  {https://doi.org/10.1088/1367-2630/16/1/013009} {\bibfield  {journal}
  {\bibinfo  {journal} {New Journal of Physics}\ }\textbf {\bibinfo {volume}
  {16}},\ \bibinfo {pages} {013009} (\bibinfo {year} {2014})}\BibitemShut
  {NoStop}%
\bibitem [{\citenamefont {Mari}\ and\ \citenamefont
  {Eisert}(2012)}]{marieisert}%
  \BibitemOpen
  \bibfield  {author} {\bibinfo {author} {\bibfnamefont {A.}~\bibnamefont
  {Mari}}\ and\ \bibinfo {author} {\bibfnamefont {J.}~\bibnamefont {Eisert}},\
  }\bibfield  {title} {\bibinfo {title} {Positive wigner functions render
  classical simulation of quantum computation efficient},\ }\href
  {https://doi.org/10.1103/PhysRevLett.109.230503} {\bibfield  {journal}
  {\bibinfo  {journal} {Phys. Rev. Lett.}\ }\textbf {\bibinfo {volume} {109}},\
  \bibinfo {pages} {230503} (\bibinfo {year} {2012})}\BibitemShut {NoStop}%
\bibitem [{\citenamefont {Nielsen}\ and\ \citenamefont
  {Chuang}(2011)}]{nielsenChuang}%
  \BibitemOpen
  \bibfield  {author} {\bibinfo {author} {\bibfnamefont {M.~A.}\ \bibnamefont
  {Nielsen}}\ and\ \bibinfo {author} {\bibfnamefont {I.~L.}\ \bibnamefont
  {Chuang}},\ }\href@noop {} {\emph {\bibinfo {title} {Quantum Computation and
  Quantum Information: 10th Anniversary Edition}}},\ \bibinfo {edition} {10th}\
  ed.\ (\bibinfo  {publisher} {Cambridge University Press},\ \bibinfo {address}
  {New York, NY, USA},\ \bibinfo {year} {2011})\BibitemShut {NoStop}%
\bibitem [{Note2()}]{Note2}%
  \BibitemOpen
  \bibinfo {note} {See Ref.~\cite { campbell2017roads}. For a reminder, see
  Appendix \ref {app:reminder}.}\BibitemShut {Stop}%
\bibitem [{Note4()}]{Note4}%
  \BibitemOpen
  \bibinfo {note} {Perhaps one cannot directly sample $\eta _x$. In fact,
  according to the definition of $\protect \mathcal {Q}_t(\sigma _r)$, we may
  be able to only prepare some states $\phi _i$ such that $\eta _x = \DOTSB
  \sum@ \slimits@ _{i} h^{(x)}_i \phi _i$ for a probability $\{h^{(x)}_i\}_i$.
  If that's the case, sample $\phi _i$ with probability $h^{(x)}_i$. An
  application of Hoeffding's inequality shows that this sampling has the same
  overhead as the original one.}\BibitemShut {Stop}%
\bibitem [{\citenamefont {Hoeffding}(1963)}]{hoeffding1963}%
  \BibitemOpen
  \bibfield  {author} {\bibinfo {author} {\bibfnamefont {W.}~\bibnamefont
  {Hoeffding}},\ }\bibfield  {title} {\bibinfo {title} {Probability
  inequalities for sums of bounded random variables},\ }\href
  {https://doi.org/10.1080/01621459.1963.10500830} {\bibfield  {journal}
  {\bibinfo  {journal} {Journal of the American Statistical Association}\
  }\textbf {\bibinfo {volume} {58}},\ \bibinfo {pages} {13} (\bibinfo {year}
  {1963})}\BibitemShut {NoStop}%
\bibitem [{\citenamefont {Boyd}\ and\ \citenamefont
  {Vandenberghe}(2004)}]{boyd}%
  \BibitemOpen
  \bibfield  {author} {\bibinfo {author} {\bibfnamefont {S.}~\bibnamefont
  {Boyd}}\ and\ \bibinfo {author} {\bibfnamefont {L.}~\bibnamefont
  {Vandenberghe}},\ }\href@noop {} {\emph {\bibinfo {title} {Convex
  Optimization}}}\ (\bibinfo  {publisher} {Cambridge University Press},\
  \bibinfo {address} {USA},\ \bibinfo {year} {2004})\BibitemShut {NoStop}%
\bibitem [{Note5()}]{Note5}%
  \BibitemOpen
  \bibinfo {note} {It is possible to alleviate this problem by exploiting the
  symmetries of the state $\tau ^{\otimes t}$ \cite
  {heinrichGross2019}.}\BibitemShut {Stop}%
\bibitem [{\citenamefont {Seddon}\ \emph {et~al.}(2021)\citenamefont {Seddon},
  \citenamefont {Regula}, \citenamefont {Pashayan}, \citenamefont {Ouyang},\
  and\ \citenamefont {Campbell}}]{seddon2020}%
  \BibitemOpen
  \bibfield  {author} {\bibinfo {author} {\bibfnamefont {J.~R.}\ \bibnamefont
  {Seddon}}, \bibinfo {author} {\bibfnamefont {B.}~\bibnamefont {Regula}},
  \bibinfo {author} {\bibfnamefont {H.}~\bibnamefont {Pashayan}}, \bibinfo
  {author} {\bibfnamefont {Y.}~\bibnamefont {Ouyang}},\ and\ \bibinfo {author}
  {\bibfnamefont {E.~T.}\ \bibnamefont {Campbell}},\ }\bibfield  {title}
  {\bibinfo {title} {Quantifying quantum speedups: Improved classical
  simulation from tighter magic monotones},\ }\href
  {https://doi.org/10.1103/PRXQuantum.2.010345} {\bibfield  {journal} {\bibinfo
   {journal} {PRX Quantum}\ }\textbf {\bibinfo {volume} {2}},\ \bibinfo {pages}
  {010345} (\bibinfo {year} {2021})}\BibitemShut {NoStop}%
\bibitem [{Note6()}]{Note6}%
  \BibitemOpen
  \bibinfo {note} {For example, the $k=3$ block decomposition error mitigates a
  quantum circuit with depolarizing noise $\delta =10^{-2}$ on each $T$-state
  with sampling overhead $1.0201^t$, while $k=1$ achieves $1.0203^t$. Formally,
  we conjecture $\protect \mathcal {R}(\tau ^{\otimes t} | \tau _\delta
  ^{\otimes t}) \approx (1 + \delta )^t$ as $\delta \rightarrow
  0$.}\BibitemShut {Stop}%
\bibitem [{\citenamefont {Bravyi}\ and\ \citenamefont
  {Gosset}(2016)}]{bravyigosset2016}%
  \BibitemOpen
  \bibfield  {author} {\bibinfo {author} {\bibfnamefont {S.}~\bibnamefont
  {Bravyi}}\ and\ \bibinfo {author} {\bibfnamefont {D.}~\bibnamefont
  {Gosset}},\ }\bibfield  {title} {\bibinfo {title} {Improved {C}lassical
  {S}imulation of {Q}uantum {C}ircuits {D}ominated by {C}lifford {G}ates},\
  }\href {https://doi.org/10.1103/PhysRevLett.116.250501} {\bibfield  {journal}
  {\bibinfo  {journal} {Phys. Rev. Lett.}\ }\textbf {\bibinfo {volume} {116}},\
  \bibinfo {pages} {250501} (\bibinfo {year} {2016})}\BibitemShut {NoStop}%
\bibitem [{\citenamefont {{Pashayan}}\ \emph {et~al.}(2021)\citenamefont
  {{Pashayan}}, \citenamefont {{Reardon-Smith}}, \citenamefont {{Korzekwa}},\
  and\ \citenamefont {{Bartlett}}}]{pashayankorzekwa2021}%
  \BibitemOpen
  \bibfield  {author} {\bibinfo {author} {\bibfnamefont {H.}~\bibnamefont
  {{Pashayan}}}, \bibinfo {author} {\bibfnamefont {O.}~\bibnamefont
  {{Reardon-Smith}}}, \bibinfo {author} {\bibfnamefont {K.}~\bibnamefont
  {{Korzekwa}}},\ and\ \bibinfo {author} {\bibfnamefont {S.~D.}\ \bibnamefont
  {{Bartlett}}},\ }\href@noop {} {\bibinfo {title} {{Fast estimation of outcome
  probabilities for quantum circuits}}} (\bibinfo {year} {2021}),\ \Eprint
  {https://arxiv.org/abs/2101.12223} {arXiv:2101.12223 [quant-ph]} \BibitemShut
  {NoStop}%
\bibitem [{Note7()}]{Note7}%
  \BibitemOpen
  \bibinfo {note} {For the sake of comparison, setting $\delta =10^{-2}$ and
  \protect \mbox {$\delta _c = 10^{-3}$} we get an overhead $1.02845^t \times
  1.00301^{n^{(1)}_c} \times 1.00376^{n^{(2)}_c}$.}\BibitemShut {Stop}%
\bibitem [{\citenamefont {Yoganathan}\ \emph {et~al.}(2019)\citenamefont
  {Yoganathan}, \citenamefont {Jozsa},\ and\ \citenamefont
  {Strelchuk}}]{yoganathan2019quantum}%
  \BibitemOpen
  \bibfield  {author} {\bibinfo {author} {\bibfnamefont {M.}~\bibnamefont
  {Yoganathan}}, \bibinfo {author} {\bibfnamefont {R.}~\bibnamefont {Jozsa}},\
  and\ \bibinfo {author} {\bibfnamefont {S.}~\bibnamefont {Strelchuk}},\
  }\bibfield  {title} {\bibinfo {title} {Quantum advantage of unitary clifford
  circuits with magic state inputs},\ }\href@noop {} {\bibfield  {journal}
  {\bibinfo  {journal} {Proceedings of the Royal Society A}\ }\textbf {\bibinfo
  {volume} {475}},\ \bibinfo {pages} {20180427} (\bibinfo {year}
  {2019})}\BibitemShut {NoStop}%
\bibitem [{\citenamefont {Bravyi}\ \emph
  {et~al.}(2016{\natexlab{a}})\citenamefont {Bravyi}, \citenamefont {Smith},\
  and\ \citenamefont {Smolin}}]{bravyi2016trading}%
  \BibitemOpen
  \bibfield  {author} {\bibinfo {author} {\bibfnamefont {S.}~\bibnamefont
  {Bravyi}}, \bibinfo {author} {\bibfnamefont {G.}~\bibnamefont {Smith}},\ and\
  \bibinfo {author} {\bibfnamefont {J.~A.}\ \bibnamefont {Smolin}},\ }\bibfield
   {title} {\bibinfo {title} {Trading classical and quantum computational
  resources},\ }\href {https://doi.org/10.1103/PhysRevX.6.021043} {\bibfield
  {journal} {\bibinfo  {journal} {Phys. Rev. X}\ }\textbf {\bibinfo {volume}
  {6}},\ \bibinfo {pages} {021043} (\bibinfo {year}
  {2016}{\natexlab{a}})}\BibitemShut {NoStop}%
\bibitem [{\citenamefont {Bravyi}\ \emph
  {et~al.}(2016{\natexlab{b}})\citenamefont {Bravyi}, \citenamefont {Smith},\
  and\ \citenamefont {Smolin}}]{bravyismolin2016}%
  \BibitemOpen
  \bibfield  {author} {\bibinfo {author} {\bibfnamefont {S.}~\bibnamefont
  {Bravyi}}, \bibinfo {author} {\bibfnamefont {G.}~\bibnamefont {Smith}},\ and\
  \bibinfo {author} {\bibfnamefont {J.~A.}\ \bibnamefont {Smolin}},\ }\bibfield
   {title} {\bibinfo {title} {Trading classical and quantum computational
  resources},\ }\href {https://doi.org/10.1103/PhysRevX.6.021043} {\bibfield
  {journal} {\bibinfo  {journal} {Phys. Rev. X}\ }\textbf {\bibinfo {volume}
  {6}},\ \bibinfo {pages} {021043} (\bibinfo {year}
  {2016}{\natexlab{b}})}\BibitemShut {NoStop}%
\bibitem [{\citenamefont {Bravyi}\ \emph {et~al.}(2019)\citenamefont {Bravyi},
  \citenamefont {Browne}, \citenamefont {Calpin}, \citenamefont {Campbell},
  \citenamefont {Gosset},\ and\ \citenamefont {Howard}}]{bravyicampbell2019}%
  \BibitemOpen
  \bibfield  {author} {\bibinfo {author} {\bibfnamefont {S.}~\bibnamefont
  {Bravyi}}, \bibinfo {author} {\bibfnamefont {D.}~\bibnamefont {Browne}},
  \bibinfo {author} {\bibfnamefont {P.}~\bibnamefont {Calpin}}, \bibinfo
  {author} {\bibfnamefont {E.}~\bibnamefont {Campbell}}, \bibinfo {author}
  {\bibfnamefont {D.}~\bibnamefont {Gosset}},\ and\ \bibinfo {author}
  {\bibfnamefont {M.}~\bibnamefont {Howard}},\ }\bibfield  {title} {\bibinfo
  {title} {Simulation of quantum circuits by low-rank stabilizer
  decompositions},\ }\href {https://doi.org/10.22331/q-2019-09-02-181}
  {\bibfield  {journal} {\bibinfo  {journal} {{Quantum}}\ }\textbf {\bibinfo
  {volume} {3}},\ \bibinfo {pages} {181} (\bibinfo {year} {2019})}\BibitemShut
  {NoStop}%
\bibitem [{\citenamefont {Raussendorf}\ \emph {et~al.}(2020)\citenamefont
  {Raussendorf}, \citenamefont {Bermejo-Vega}, \citenamefont {Tyhurst},
  \citenamefont {Okay},\ and\ \citenamefont {Zurel}}]{raussendorf2020}%
  \BibitemOpen
  \bibfield  {author} {\bibinfo {author} {\bibfnamefont {R.}~\bibnamefont
  {Raussendorf}}, \bibinfo {author} {\bibfnamefont {J.}~\bibnamefont
  {Bermejo-Vega}}, \bibinfo {author} {\bibfnamefont {E.}~\bibnamefont
  {Tyhurst}}, \bibinfo {author} {\bibfnamefont {C.}~\bibnamefont {Okay}},\ and\
  \bibinfo {author} {\bibfnamefont {M.}~\bibnamefont {Zurel}},\ }\bibfield
  {title} {\bibinfo {title} {Phase-space-simulation method for quantum
  computation with magic states on qubits},\ }\href
  {https://doi.org/10.1103/PhysRevA.101.012350} {\bibfield  {journal} {\bibinfo
   {journal} {Phys. Rev. A}\ }\textbf {\bibinfo {volume} {101}},\ \bibinfo
  {pages} {012350} (\bibinfo {year} {2020})}\BibitemShut {NoStop}%
\bibitem [{\citenamefont {Pashayan}\ \emph {et~al.}(2015)\citenamefont
  {Pashayan}, \citenamefont {Wallman},\ and\ \citenamefont
  {Bartlett}}]{pashayan2015}%
  \BibitemOpen
  \bibfield  {author} {\bibinfo {author} {\bibfnamefont {H.}~\bibnamefont
  {Pashayan}}, \bibinfo {author} {\bibfnamefont {J.~J.}\ \bibnamefont
  {Wallman}},\ and\ \bibinfo {author} {\bibfnamefont {S.~D.}\ \bibnamefont
  {Bartlett}},\ }\bibfield  {title} {\bibinfo {title} {Estimating outcome
  probabilities of quantum circuits using quasiprobabilities},\ }\href
  {https://doi.org/10.1103/PhysRevLett.115.070501} {\bibfield  {journal}
  {\bibinfo  {journal} {Phys. Rev. Lett.}\ }\textbf {\bibinfo {volume} {115}},\
  \bibinfo {pages} {070501} (\bibinfo {year} {2015})}\BibitemShut {NoStop}%
\bibitem [{\citenamefont {Piveteau}\ \emph {et~al.}(2021)\citenamefont
  {Piveteau}, \citenamefont {Sutter}, \citenamefont {Bravyi}, \citenamefont
  {Gambetta},\ and\ \citenamefont {Temme}}]{piveteau2021error}%
  \BibitemOpen
  \bibfield  {author} {\bibinfo {author} {\bibfnamefont {C.}~\bibnamefont
  {Piveteau}}, \bibinfo {author} {\bibfnamefont {D.}~\bibnamefont {Sutter}},
  \bibinfo {author} {\bibfnamefont {S.}~\bibnamefont {Bravyi}}, \bibinfo
  {author} {\bibfnamefont {J.~M.}\ \bibnamefont {Gambetta}},\ and\ \bibinfo
  {author} {\bibfnamefont {K.}~\bibnamefont {Temme}},\ }\bibfield  {title}
  {\bibinfo {title} {Error mitigation for universal gates on encoded qubits},\
  }\href {https://doi.org/10.1103/PhysRevLett.127.200505} {\bibfield  {journal}
  {\bibinfo  {journal} {Phys. Rev. Lett.}\ }\textbf {\bibinfo {volume} {127}},\
  \bibinfo {pages} {200505} (\bibinfo {year} {2021})}\BibitemShut {NoStop}%
\bibitem [{\citenamefont {Zhou}\ \emph {et~al.}(2000)\citenamefont {Zhou},
  \citenamefont {Leung},\ and\ \citenamefont {Chuang}}]{leung2000}%
  \BibitemOpen
  \bibfield  {author} {\bibinfo {author} {\bibfnamefont {X.}~\bibnamefont
  {Zhou}}, \bibinfo {author} {\bibfnamefont {D.~W.}\ \bibnamefont {Leung}},\
  and\ \bibinfo {author} {\bibfnamefont {I.~L.}\ \bibnamefont {Chuang}},\
  }\bibfield  {title} {\bibinfo {title} {Methodology for quantum logic gate
  construction},\ }\href {https://doi.org/10.1103/PhysRevA.62.052316}
  {\bibfield  {journal} {\bibinfo  {journal} {Phys. Rev. A}\ }\textbf {\bibinfo
  {volume} {62}},\ \bibinfo {pages} {052316} (\bibinfo {year}
  {2000})}\BibitemShut {NoStop}%
\bibitem [{\citenamefont {Gottesman}(1997)}]{gottesmanThesis}%
  \BibitemOpen
  \bibfield  {author} {\bibinfo {author} {\bibfnamefont {D.}~\bibnamefont
  {Gottesman}},\ }\emph {\bibinfo {title} {Stabilizer {C}odes and {Q}uantum
  {E}rror {C}orrection}},\ \href@noop {} {Ph.D. thesis},\ \bibinfo  {school}
  {CalTech} (\bibinfo {year} {1997})\BibitemShut {NoStop}%
\bibitem [{\citenamefont {{Suzuki}}\ \emph {et~al.}(2020)\citenamefont
  {{Suzuki}}, \citenamefont {{Endo}}, \citenamefont {{Fujii}},\ and\
  \citenamefont {{Tokunaga}}}]{suzuki2020quantum}%
  \BibitemOpen
  \bibfield  {author} {\bibinfo {author} {\bibfnamefont {Y.}~\bibnamefont
  {{Suzuki}}}, \bibinfo {author} {\bibfnamefont {S.}~\bibnamefont {{Endo}}},
  \bibinfo {author} {\bibfnamefont {K.}~\bibnamefont {{Fujii}}},\ and\ \bibinfo
  {author} {\bibfnamefont {Y.}~\bibnamefont {{Tokunaga}}},\ }\href@noop {}
  {\bibinfo {title} {Quantum error mitigation for fault-tolerant quantum
  computing}} (\bibinfo {year} {2020}),\ \Eprint
  {https://arxiv.org/abs/2010.03887} {arXiv:2010.03887 [quant-ph]} \BibitemShut
  {NoStop}%
\end{thebibliography}%

\end{document}